\documentclass[fleqn,usenatbib]{mnras}

\usepackage{newtxtext,newtxmath}

\usepackage[T1]{fontenc}

\DeclareRobustCommand{\VAN}[3]{#2}
\let\VANthebibliography\thebibliography
\def\thebibliography{\DeclareRobustCommand{\VAN}[3]{##3}\VANthebibliography}

\usepackage{lipsum}
\usepackage{mwe}
\DeclareUnicodeCharacter{2212}{-}

\usepackage{tabulary}

\usepackage{xcolor}

\usepackage{graphicx}	
\usepackage{amsmath}	
\usepackage{hyperref}
\hypersetup{
    colorlinks=true,
    linkcolor=blue,
    filecolor=magenta,      
    urlcolor=cyan,
}

\newcommand{\sigv}{\langle \sigma v \rangle}
\newcommand{\Msun}{$M_{\odot}$}

\newcommand{\kpc}{{\rm kpc}}

\newcommand{\kms}{{\rm km \, s}^{-1}}

\newcommand{\mi}{\texttt{M12i}}
\newcommand{\midark}{\texttt{M12iDMO}}

\newcommand{\mc}{\texttt{M12c}}
\newcommand{\mcdark}{\texttt{M12cDMO}}

\newcommand{\mm}{\texttt{M12m}}
\newcommand{\mmdark}{\texttt{M12mDMO}}

\newcommand{\mf}{\texttt{M12f}}
\newcommand{\mfdark}{\texttt{M12fDMO}}

\newcommand{\mw}{\texttt{M12w}}
\newcommand{\mwdark}{\texttt{M12wDMO}}

\newcommand{\mb}{\texttt{M12b}}
\newcommand{\mbdark}{\texttt{M12bDMO}}

\newcommand{\Julietdark}{\texttt{JulietDMO}}
\newcommand{\Juliet}{\texttt{Juliet}}

\newcommand{\Romeodark}{\texttt{RomeoDMO}}
\newcommand{\Romeo}{\texttt{Romeo}}

\newcommand{\Thelmadark}{\texttt{ThelmaDMO}}
\newcommand{\Thelma}{\texttt{Thelma}}

\newcommand{\Louisedark}{\texttt{LouiseDMO}}
\newcommand{\Louise}{\texttt{Louise}}

\newcommand{\Romulusdark}{\texttt{RomulusDMO}}
\newcommand{\Romulus}{\texttt{Romulus}}

\newcommand{\Remusdark}{\texttt{RemusDMO}}
\newcommand{\Remus}{\texttt{Remus}}

\newcommand{\Junit}{GeV$^2$ cm$^{-3}$}
\newcommand{\dd}{{\rm{d}}}

\newcommand{\spacer}{\hspace{.1cm} \vline \hspace{.1cm} }

\title[Galactic J-Factors from FIRE]{ Amplified J-factors in the Galactic Center for velocity-dependent dark matter annihilation in FIRE simulations  }

\author[Daniel McKeown et al.]
{\parbox{17.5cm}{
Daniel McKeown$^{1}$\thanks{E-mail: dmckeown@uci.edu}, James S. Bullock$^{1}$, Francisco J. Mercado $^{1}$, Zachary Hafen$^{1}$, Michael Boylan-Kolchin$^{2}$, Andrew Wetzel$^{3}$, Lina Necib$^{4}$, Philip F. Hopkins$^{4}$, Sijie Yu$^{1}$ } \vspace{0.3cm}\\
$^{1}$Center for Cosmology, Department of Physics and Astronomy,4129 Reines Hall, University of California Irvine, CA 92697, USA \\
$^{2}$Department of Astronomy, The University of Texas at Austin, 2515 Speedway Stop C1400, Austin, TX 78712, USA\\
$^{3}$Department of Physics, University of California, Davis, CA 95616, USA\\
$^{4}$TAPIR, Mailcode 350-17, California Institute of Technology, Pasadena, CA 91125, USA\\
}

\date{Accepted XXX. Received YYY; in original form ZZZ}

\pubyear{2020}

\date{Accepted XXX. Received YYY; in original form ZZZ}

\pubyear{2021}

\begin{document}
\label{firstpage}
\pagerange{\pageref{firstpage}--\pageref{lastpage}}
\maketitle

\begin{abstract}
We use FIRE-2 zoom cosmological simulations of Milky Way size galaxy halos to calculate astrophysical J-factors for dark matter annihilation and indirect detection studies. In addition to velocity-independent (s-wave) annihilation cross sections $\sigv$, we also calculate effective J-factors for velocity-dependent models, where the annihilation cross section is either either p-wave ($\propto v^2/c^2$) or d-wave ($\propto v^4/c^4$). We use 12 pairs of simulations, each run with dark-matter-only (DMO) physics and FIRE-2 physics. We observe FIRE runs produce central dark matter velocity dispersions that are systematically larger than in DMO runs by factors of $\sim 2.5-4$. They also have a larger range of central ($\sim 400$ pc) dark matter densities than the DMO runs ($\rho_{\rm FIRE}/\rho_{\rm DMO} \simeq 0.5 - 3$) owing to the competing effects of baryonic contraction and feedback. At 3 degrees from the Galactic Center, FIRE J-factors are $3-60$ (p-wave) and $10-500$ (d-wave) times higher than in the DMO runs. The change in s-wave signal at 3 degrees is more modest and can be higher or lower ($\sim 0.3-7$), though the shape of the emission profile is flatter (less peaked towards the Galactic Center) and more circular on the sky in FIRE runs. Our results for s-wave are broadly consistent with the range of assumptions in most indirect detection studies. We observe p-wave J-factors that are significantly enhanced compared to most past estimates. We find that thermal models with p-wave annihilation may be within range of detection in the near future.

\end{abstract}




\section{Introduction}

There is significant astrophysical and cosmological evidence showing that non-baryonic dark matter dominates the mass in the Universe \citep[e.g.][]{Einasto74,Rubin78,Trimble87,Wittman2000,Mandelbaum2013,Kwan2017}.  Cosmological constraints have pinpointed that the dark matter mass density relative to critical is $\Omega_{\rm dm} \simeq 0.27$ today \citep[see][and references therein]{Planck2020}. For a comprehensive historical perspective on the observational and theoretical motivations for dark matter see \citet{BH18}.

One popular theory for dark matter suggests that it is made up of Weakly Interacting Massive Particles (WIMPs) \citep[WIMPs, see][for discussions]{Jungman96,dodelson2003modern,Feng05,Bertone05,mo2010galaxy}. In the standard WIMP scenario, where dark matter particles are their own antiparticles, WIMPs self-annihilate and recombine in equilibrium when the Universe is young, hot, and dense. As the Universe cools and expands, annihilation rates become too low to maintain equilibrium, and the co-moving particle abundance “freezes out.” The resultant abundance is set directly by the interaction cross section during freeze-out, and this gives us a way to relate a macroscopic observable ($\Omega_{\rm dm} \simeq  0.27$) to microscopic properties of the particle.  Specifically, if the thermally-averaged cross-section during freeze out is $\langle \sigma_A v \rangle \equiv \langle \sigma_A v \rangle_T \simeq 2.3 \times 10^{-26}$ cm$^3$ s$^{-1}$, then the thermal relic density is naturally of the right order of magnitude to match the observed abundance \citep[see][for a precise treatment]{Steigman12}.

The same annihilations that set the thermal abundance of WIMPs in the early Universe should be occurring again today in regions where the dark matter has become dense in dark matter halos.  If those annihilations produce Standard Model particles, this provides a means for indirect dark matter detection.  Specifically, an observed flux of Standard Model particles from such a location could provide evidence for dark matter.
One region of particular interest is the Galactic Center \citep[][]{Bergstrom88}.  Not only is the Galactic Center expected to be dense in dark matter but its relative proximity to Earth has made it a subject of significant study for indirect messengers of annihilation, including cosmic rays and neutrinos \citep[see][for a review]{Gaskins2016}.   If annihilation to quarks and charged lepton states happens for dark matter particles, this will ultimately produce photons with energies of order $\sim 10\%$ of the dark matter particle mass, making gamma-ray observations of particular interest for indirect searches for WIMP dark matter with $m_\chi \sim 100$ GeV \citep{daylan2016characterization}.

An observed excess in gamma-ray emission from the Galactic Center based on Fermi Large Area Telescope observations has sparked considerable interest as a potential indirect detection signal \citep{Hooper11,Abazajian12,martin2014fitting,Ajello16,Ackerman17}. 
The basic excess has been confirmed by multiple groups \citep[see][for a review]{murgia2020} and is consistent with expectations for a dark matter particle with mass $m_\chi \sim 10-100$ GeV annihilating with a velocity-averaged cross section that matches the thermal WIMP expectation.  A different signal from the Andromeda galaxy halo is potentially consistent with this interpretation \citep{Karwin2020}.    
Although a dark matter origin of the Galactic Center Excess (GCE) is the most intriguing possibility, astrophysical sources, including gamma-ray emitting pulsars \citep[e.g.][]{Abazajian11,Bartels16} and supernovae remnants \citep[e.g.][]{carlson2014cosmic} are plausible alternatives. 

The case for an astrophysical interpretation has been strengthened over the last several years, with analyses showing that the morphology of the excess traces the flattened "boxy" stellar over-density of the Galactic bulge, rather than the more spherical distribution one would expect for a dark matter signal \citep{Macias18,Bartels18}.  Based on this realization, \citet{Abazajian20} used templates for the galactic and nuclear stellar bulges to show that the GC shows no significant evidence for DM annihilation and used this to place strong constraints on the s-wave cross section.  In particular, in the case of a pure b-quark annihilation channel, assuming a range of DM profiles consistent with numerical simulations, \citet{Abazajian20} ruled out s-wave cross sections $ \gtrsim 0.015 \langle \sigma_A v \rangle_T$ for dark matter masses $m_\chi \sim 10$ GeV.

One way that this thermal limit could be avoided is if dark matter annihilation is velocity-dependent 
\citep{Robertson_Zentner09,Giacchino13,Choquette16,Petac18,Boddy18,johnson2019search,Arguelles19,Board21}.  Specifically, in some models, symmetries forbid the s-wave contribution to the annihilation cross-section, and the leading contribution to DM annihilations could be p-wave $\sigma v \propto (v/c)^2$ or even d-wave $\sigma v \propto (v/c)^4$.  In the Milky Way, typical DM velocities are usually thought to be $\sim 3 \times 10^{-4} c$ near the Galactic Center, while at thermal freeze out 
$v \sim 0.1 c$.  This means that for p-wave, the cross section is expected to be suppressed by a factor of $\sim 10^{-5}$ compared to the value during freeze-out.  

For a fixed particle physics model with s-wave annihilation, the expected annihilation signal depends on the square of the dark matter density along the line-of-sight from the observer.  This "astrophysical J-factor" is therefore critical to the interpretation of any indirect dark matter search \citep{Stoehr03,Diemand07,Springel08,Kamionkowski10,Grand21}.  It is common for interpretive analyses to adopt analytic priors for the dark matter profile shape inferred from  cosmological simulations and to normalize the profiles so that the local dark matter density near the Sun matches observationally-inferred values \citep{necib2019under}. For velocity-dependent models, the J-factor is generalized to include the local velocity distribution \citep{Boddy18}. Recently, \citet{Board21} used several cosmological zoom hydrodynamical simulations to investigate the generalized J-factor for velocity dependent models for Milky-Way size galaxies.  They found that J-factors were enhanced for hydrodynamic runs in p and d-wave cases.  They also concluded that the J-factor in all models was strongly correlated with the local dark matter density.

In this paper, we perform a similar analysis to that in \citet{Board21} utilizing 12 Milky Way mass zoom simulations done as part of the FIRE-2 collaboration \citep{Wetzel16,garrison2017not,Hopkins17,Garrison-Kimmel19,lazar2020dark}. For each halo, we have a dark-matter-only version, and this allows us to explore the differential effect of galaxy formation physics on J-factor predictions. 

Our work extends that of \citet{Board21} in three significant ways.  First, our simulations have $\sim 10$ times better mass resolution and this allows us to resolve the J-factor $\sim 3$ times closer the Galactic Center (within 2.75$^ {\circ}$) than they were able to do.  Second, we do not assume spherical symmetry in our analysis, and this allows us to explore the shape of emission on the sky. Finally, we re-normalize every halo so that the local dark matter density at mock solar locations are identical.  This allows us to mimic what is done in indirect detection analyses and to explore how differences in the shape of the dark matter density and velocity profile will affect J-factor predictions from simulation to simulation in a way scales out the expected dependence on local density.

The outline of this paper is as follows.  In section 2 we describe our simulations.  In section 3 we provide our nomenclature for the effective J-factor and describe our analysis. In section 4 present our results and conclude in section 5.

\begin{table*}
	\centering
	\caption{(1) Simulation name. The suffix ``DMO" stands for ``Dark Matter Only" and refers to the same simulation run with no hydrodynamics or galaxy formation physics. (2) Factor $f$ by which dark matter particle masses have been multiplied ($m_{\rm dm} \rightarrow fm_{\rm dm}$) in order to normalize the dark matter density at $\rho(r = R_\odot) = 10^7$ \Msun  $\kpc^{-3} =  0.38 $GeV cm$^{-3}$ for a mock solar location $R_\odot = 8.3$ kpc. (3) Stellar mass $M_\star$ of the central galaxy. (4) Virial mass (of raw simulation, not including the $f$ factor) defined by \citet{ByranNorman1998}. The following quantities are derived after normalizing (by $f$) to the local dark matter density: (5) Cumulative s-wave J-factor within 3 degrees of the Galactic Center.  (6) Cumulative s-wave J-factor within 10 degrees of the Galactic Center. (7) Cumulative s-wave J-factor integrated over the sky. (8) Cumulative p-wave J-factor within 3 degrees of the Galactic Center.  (9) Cumulative p-wave J-factor within 10 degrees of the Galactic Center. (10) Cumulative p-wave J-factor integrated over the sky. (11) Cumulative d-wave J-factor within 3 degrees of the Galactic Center.  (12) Cumulative d-wave J-factor within 10 degrees of the Galactic Center. (13) Cumulative d-wave J-factor integrated over the sky.
	}
	\label{tab:one}
	\begin{tabular}{lccc @{\spacer} ccc @{\spacer} ccc @{\spacer} ccc} 
		\hline
		Simulation & f & M$_\star$ & M$_{\rm vir}$ & J$_s$(< $3^{\circ}$) & J$_s$(< $10^{\circ}$) & J$_s^{\rm tot}$ & J$_p$(< $3^{\circ}$)  & J$_p$(< $10^{\circ}$) & J$_p^{\rm tot}$ & J$_d$(< $3^{\circ}$) & J$_d$(< $10^{\circ}$) & J$_d^{\rm tot}$ \\ 
		           &   & $10^{10}$ M$_\odot$ &  $10^{12}$ M$_\odot$ & 
		             &  $\hspace{-2 cm}$ ($10^{22}$ & \hspace{-2 cm}   \Junit) & &  \hspace{-2 cm} ($10^{16}$ & \hspace{-2 cm}   \Junit) &  &  \hspace{-2 cm} ($10^{10}$ & \hspace{-2 cm}   \Junit) \\  
		\hline
		\mi         & 1.28 & 6.4  & 0.90  & 1.34 & 7.05 & 17.6 & 4.90 & 22.1 & 44.7 & 24.6 & 95.9 & 163 \\
		\midark     & 1.59 & -   & 1.3  & 0.489 & 2.33 & 9.58 & 0.137 & 0.913 & 5.73 & 0.066 & 0.572 & 5.13 \\ \\
		
		\mc         & 1.26  & 6.0  & 1.1  & 1.16 & 6.27 & 17.6 & 3.74 & 17.8 & 39.8 & 16.6 & 69.6 & 130  \\
		\mcdark     & 1.83  & -    & 1.3  & 1.27 & 5.10 & 14.3 & 0.294 & 1.65 & 6.62 & 0.112 & 0.843 & 4.70 \\ \\
		
		\mm         &  0.885 &  11 & 1.2  & 0.607 & 3.92 & 13.9 & 2.46 & 14.3 & 41.3 & 13.7 & 71.9 &  174 \\
		\mmdark    &  1.42  &  - & 1.4 & 1.56 & 6.12 &  15.3 & 0.490 & 2.52 & 8.56 & 0.237 & 1.56 & 7.21  \\ \\
		
		\mf         &  1.01  &  8.6 & 1.3 &  0.978 & 5.93 & 16.2 & 4.05 & 21.3 & 47.2 & 23.2 & 106 & 197   \\
		\mfdark    &  1.82  &  - &  1.6 &  1.22 & 4.87 & 14.1 & 0.316 & 1.73 &  7.65 & 0.131 & 0.960 & 6.54 \\ \\
		
		\mw        &  1.28  &  5.8  &  0.83  & 1.32 & 5.60 & 15.6 & 3.91 & 15.5 & 34.8 & 15.9 & 58.6 & 111   \\
		\mwdark    &  1.68  &  - & 1.1 &   0.798  & 3.24 & 11.0 & 0.212 & 1.26 & 6.23 & 0.093 & 0.757 & 5.18 \\ \\
		
		\mb          &  0.990 &  8.1  &  1.1  & 1.17 & 6.97 & 17.7 & 6.49 & 31.2 & 60.5 & 50.6 & 198 & 310 \\
		\mbdark      &  1.25  &  -  & 1.4 & 2.06 & 7.09 & 18.6 & 0.791 & 3.74 & 13.6 & 0.462 & 2.96 & 14.7  \\  \\

		\Romeo      & 0.99  & 7.4  & 1.0 & 4.50 & 15.7 & 27.7 & 10.9 & 37.7 & 64.2 & 36.1 & 123 & 203   \\
		\Romeodark  & 1.26  & - & 1.2  & 3.07 & 8.83 & 18.3 &  1.08 & 4.30 & 12.5 & 0.584 & 3.22 & 13.1 \\ \\
	
			\Juliet     & 1.31  & 4.2  & 0.85  & 4.43 & 15.5 & 26.4 & 11.4 & 35.0 & 52.4 & 40.6 &  109 & 148 \\
		\Julietdark & 1.59  & -    &  1.0   & 3.72 & 11.0 & 23.7 & 1.13 & 4.74 & 12.6 & 0.543 & 3.08 & 10.0  \\ \\

		\Thelma      & 1.17 & 7.9 & 1.1 & 0.391 & 2.81 & 11.5  & 1.31 & 8.67 &  28.5  & 6.04 & 36.7 & 99.9 \\
		\Thelmadark  & 1.70  & -    & 1.3  &  1.27 & 4.80 & 13.2 & 0.310 & 1.69 & 7.36 &  0.129 & 0.986 & 6.55 \\  \\

		\Louise     &  1.42  & 2.9    & 0.85  & 1.28  & 6.85  & 17.9 & 2.17 & 10.7 & 25.4 & 4.99 & 22.9 & 49.6 \\
		\Louisedark  & 1.41  & -    & 1.0  &  1.79 & 6.68  & 19.6 & 0.661 & 3.40 & 12.6 & 0.373 & 2.55 & 11.7 \\  \\

		\Romulus     & 1.00  & 10   & 1.53 & 7.95 & 19.9 & 31.4 & 28.5 & 66.8  &  97.8  & 139  & 306 & 421 \\
		\Romulusdark  &  1.01   & -    & 1.9  &  1.19 & 4.83 & 13.9  & 0.531 & 3.00  & 12.5  & 0.385 & 2.91 &  17.0  \\ \\

		\Remus    & 1.10  & 5.1  & 0.97 & 2.04 & 8.90 & 19.9 & 4.98 & 20.5 & 41.3 & 16.6 & 64.3 & 119 \\
		\Remusdark  & 1.18  & -  & 1.3 &  2.26 & 7.91 & 19.7 & 0.933 & 4.46 & 14.9 & 0.596 & 3.78 & 16.8 \\ \\

		\hline
	\end{tabular}
\end{table*}

\section{Overview of Simulations}

Our analysis relies on cosmological zoom-in simulations performed as part of the Feedback In Realistic Environments (FIRE) project\footnote{\url{https://fire.northwestern.edu/}} with FIRE-2 feedback implementation \citep{Hopkins17} with the gravity plus hydrodynamics code {\small GIZMO} \citep{Hopkins15}.  FIRE-2 includes radiative heating and cooling for gas with temperatures ranging from 10\,{\rm K} to $10^{10}\,{\rm K}$, an ionising background \citep{Faucher2009}, stellar feedback from OB stars, AGB mass-loss, type Ia and type II supernovae, photoelectric heating, and radiation pressure. Star formation occurs in gas that is locally self-gravitating, sufficiently dense ($ > 1000$ cm$^{-3}$), Jeans unstable, and molecular (following \citealt{Krumholz_2011}). Locally, the star formation efficiency is set to $100\%$ per free-fall time, though the global efficiency of star formation within a giant-molecular cloud (or across larger scales) is self-regulated by feedback to $\sim$1-10\% per free-fall time \citep{Orr_2018}.

In this work, we analyse 12 Milky-Way-mass galaxies (Table \ref{tab:one}). These zoom simulations are initialised following the approach outlined in \citet{Onorbe14} using the MUSIC code \citep{HA11}. Six of these galaxies were run as part of the Latte suite ~\citep{Wetzel16,Garrison-Kimmel17,Garrison-Kimmel19,samuel2020profile,Hopkins17} and have names following the convention \texttt{m12*}. The other six, with names associated with famous duos, are set in paired configurations to mimic the Milky Way and M31 \citep{Garrison-Kimmel19,Garrison-Kimmel19_2}. Analysis has shown these are good candidates for comparison with the Milky Way \citep{sanderson2020synthetic}. Gas particles for the \texttt{M12*} runs have initial masses of $m_{\rm g,i} = 7070\,$ \Msun.  The ELVIS on FIRE simulations have roughly two times better mass resolution ($m_{\rm g,i} \simeq 3500 - 4000$ \Msun). Gas softening lengths are fully adaptive down to $\simeq$0.5$-$1 pc. The dark matter particle masses are $m_{\rm dm} = 3.5 \times 10^4$ \Msun for the Latte simulations and $m_{\rm dm} \simeq 2 \times 10^4$ \Msun for the ELVIS runs.  Star particle softening lengths are $\simeq$4 pc physical and a dark matter force softening is $\simeq$40 pc physical. 

Lastly, each FIRE simulation has an analogous dark matter only (DMO) version. The individual dark matter particle masses in the DMO simulations are larger by a factor of $(1 − f_{\rm b})^{−1}$ in order to keep the total gravitating mass of the Universe the same, where  $f_{\rm b} = \Omega_{ \rm b} /\Omega_{\rm m}$ is the cosmic baryon fraction.  The initial conditions are otherwise identical. DMO versions of each halo are referred to with the same name as the FIRE version with the added suffix ``DMO."

As can be seen in Table \ref{tab:one}, the stellar masses of the main galaxy in each FIRE run (second column) are broadly in line with the Milky Way: $M_\star \approx (3 - 11) \times 10^{10}$ \Msun.  
The virial masses \citep{ByranNorman1998} of all the halos in these simulation span a range generally in line with expectations for the Milky Way: $M_{\rm vir} \approx (0.9 - 1.8) \times 10^{12} $ \Msun.   In every case, the DMO version of each pair ends up with a higher virial mass.  This is consistent with the expectation that halos will have lost their share of cosmic mass by not retaining all baryons in association with feedback. As we discuss in the next section, in our primary analysis we re-normalize all halos (both FIRE and DMO runs) so that they have the same ``local" dark matter density at the Solar location (by the factor $f$ listed in the table).

\begin{figure*}
	\includegraphics[width =0.99\columnwidth,trim = 0 0 0 0 ]{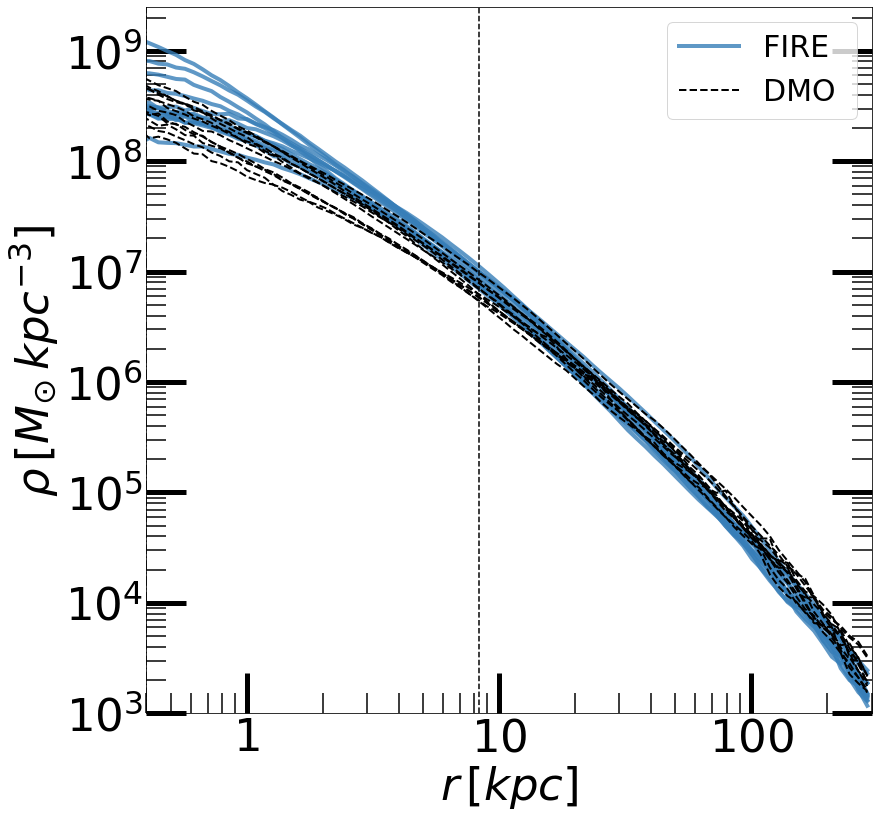} 
	\hspace{0.1in}
		\includegraphics[width = 0.99 \columnwidth, trim = 0 0 0 0]{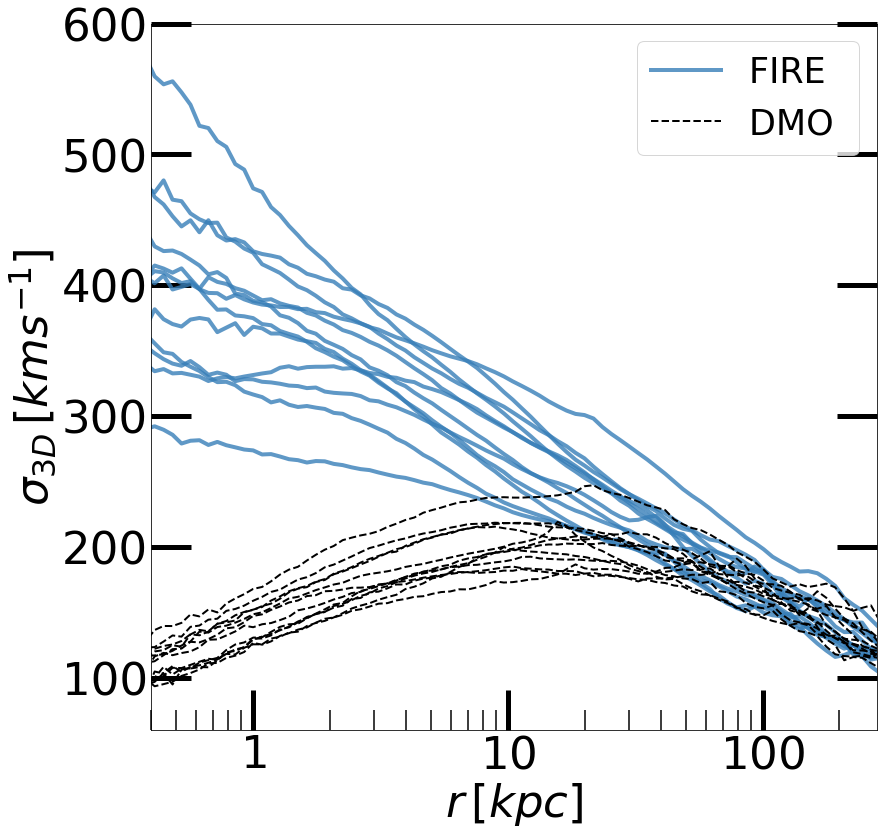} \\
		
	\includegraphics[width =0.99\columnwidth,trim = 0 0 0 0 ]{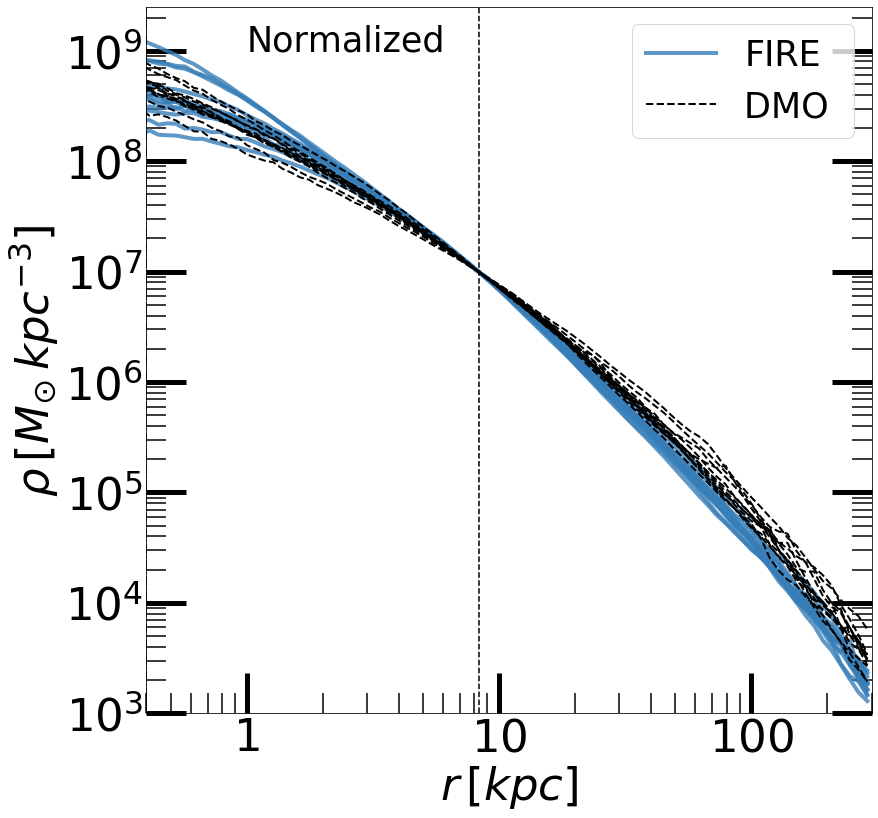} 
		\hspace{0.1in}
		\includegraphics[width =0.99\columnwidth,trim = 0 0 0 0]{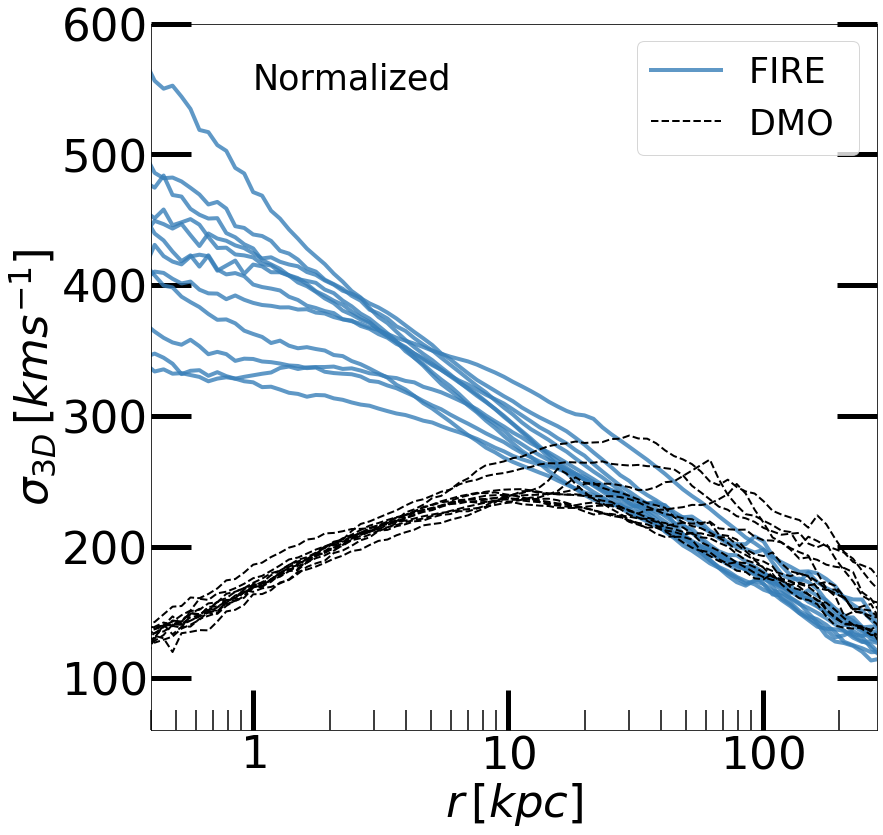}
    \caption{  {\bf Upper Left:} Simulated dark matter halo density profiles for all DMO (dashed) and FIRE (solid) runs (solid). {\bf Upper Right:} Three-dimensional dark matter velocity dispersion profiles for all DMO (dashed) and FIRE (solid) simulations.  {\bf Lower Left:}  Dark matter density profiles after re-normalizing.  The mass per particle in each simulation has been multiplied by a factor $f$ (ranging from $f=0.89 - 1.8$, see Table 1) to give them the same density at a mock solar location:  $\rho(r = R_\odot) = 10^7$ \Msun  $\kpc^{-3}$  with $R_\odot = 8.3$ kpc (vertical dotted line).   {\bf Bottom Right:} Velocity dispersion profiles for each halo after re-scaling the particle velocities by a factor $\sqrt{f}$.  This roughly accounts for the re-scaling of the mass/density profile. Note that after re-scaling, the DMO velocity dispersion profiles become similar for $r \lesssim R_\odot$. E}ven when normalized at the solar radius, there is almost an order of magnitude scatter in the inner ($\sim 400$pc) density for the FIRE simulations and they all have higher central dark matter velocities than would have been expected from DMO simulations.
    \label{fig:profiles}
\end{figure*}

\begin{figure*}
		\includegraphics[height =0.95 \columnwidth]{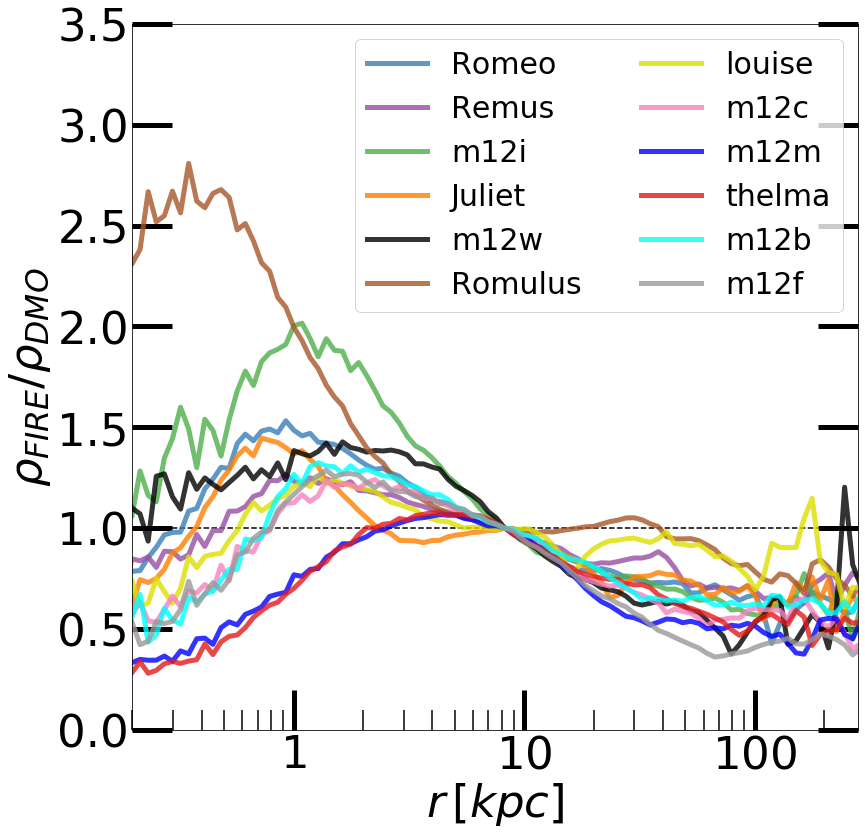}
	\hspace{.1in}
\includegraphics[height=.95\columnwidth, trim = 0 0 0 0 ]{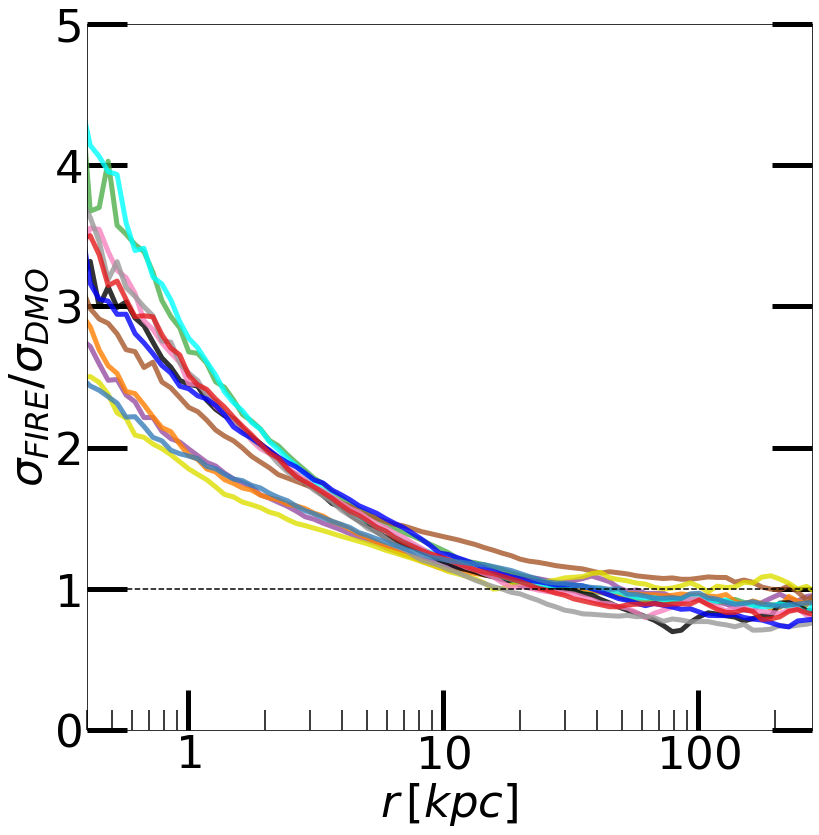}
   \caption{{\bf Left:} The ratio of the FIRE dark matter density profiles shown in Figure \ref{fig:profiles} to its DMO counterpart. We see that in some cases the FIRE runs are less dense than the DMO runs though most halos get denser at small radius in response to galaxy formation.  {\bf Right:} The ratio of the FIRE dark matter halo velocity dispersion profiles shown in Figure \ref{fig:profiles} divided by the DMO version for each halo. We see that in every case, the process of galaxy formation heats the dark matter at small radius.
   }    
    \label{fig:ratios}
\end{figure*}

\subsection{Dark Matter Density and Velocity Dispersion Profiles}

\citet{lazar2020dark} provide an extensive discussion of the dark matter halo density profiles for the simulations we analyze here. Every system has more than $\sim 1000$ dark matter particles within the inner 400pc and is converged outside of this radius according to the criteria discussed \citet{Hopkins17}.  Though some systems are even better converged, for simplicity we adopt the same convergence radius, $r_{\rm conv} = 400$pc, for each halo and only present values that depend on quantities outside of this radius. For an adopted Solar location at radius $R_\odot = 8.3$ kpc, our $400$ pc convergence radius corresponds to an angle $\psi = 2.75^\circ$ in projection at the Galactic Center.

Figure \ref{fig:profiles} shows the spherically-averaged density and velocity dispersion profiles of the simulations in our sample.  The upper two panels show raw simulation results, with differential density profiles on the left and velocity-dispersion profiles on the right.  The DMO simulations are in black and FIRE runs are in blue. Note that the density profiles of the FIRE runs are systematically steeper for $r \gtrsim 3$ kpc than the DMO runs.  This is consistent with the expectations that baryonic contraction  makes halos more concentrated at this stellar-mass scale \citep[see][]{lazar2020dark}.  At smaller radii ($r \lesssim 3$ kpc), however, the FIRE halos have a larger range of central densities; sometimes feedback produces a core-like profile and sometimes the halo remains fairly steep \citep[see][for an investigation into the origin of this variation] {mercado2021relationship}.  

The FIRE halos also have systematically higher central dark matter velocity dispersion than the DMO runs.  A similar result was reported by \citet{Robles19}, which studied the velocity dispersion profiles of cold dark matter halos using zoom in dark matter simulations that included a slowly-grown Milky-Way disk potential.  In these simulations, even without feedback, they found that the central velocity dispersion of the dark matter was much higher in runs with disk potentials compared to those without. 
\citet{Board21} also found that the central dark matter velocity dispersion was higher in simulations that included full galaxy formation physics (though with a different implementation than our own).  Taken together, these results suggest that the dark matter velocity dispersion at the center of Milky-Way mass halos should be significantly higher than would be expected from DMO simulations, irrespective of galaxy formation model.

The pair of panels on the bottom of Figure \ref{fig:profiles} show the profiles after we have re-scaled them to the defaults we will use in the rest of the analysis. Our aim here is to normalize each run to have the same local dark matter density at the solar location $r = R_\odot$. We are motivated to do this because it is customary in indirect-detection analyses to normalize the assumed profile at $R_\odot$ and to marginalize about the local density range inferred by observations.  While our halos are Milky-Way like in virial mass and stellar mass, they are not precise replicas of the Milky Way.  By re-normalizing at the solar location, our results become primarily about profile {\em shape} rather normalization, and can be scaled appropriately as observational estimates of the local density become more precise.    We assume $R_\odot = 8.3$ kpc (vertical dotted lines in the left panels) and set the density there to be $\rho(r = R_\odot) = 10^7$ \Msun  $\kpc^{-3} =  0.38 $ GeV cm$^{-3}$ \citep{Guo20}.
We do this by scaling the particle masses in each simulation (post process) by a factor $f$: m$_{\rm dm} \rightarrow f \, $m$_{\rm dm}$.  The values of $f$ for each simulation are given in Table 1 and range from $f = 1.8$ to $f=0.89$. 
We also re-scale the particle velocities in each simulation by a factor\footnote{This assumes $v \propto \sqrt{f M/r}$.  We have checked that the dark matter velocity dispersion in our simulations does roughly scale with the local dark matter density as $\sqrt{f}$.} $\sqrt{f}$ in order to roughly account for the re-scaling of the total mass: $v \rightarrow \sqrt{f} v$. Note that after re-scaling, the DMO velocity dispersion profiles become similar for $r \lesssim R_\odot$, as expected.

Even when normalized at the Solar radius, there is considerable scatter in the inner density and the FIRE simulations display more variance than the DMO simulations. The difference between DMO and FIRE is most systematic in the inner velocity dispersion (bottom right of Figure \ref{fig:profiles}).  While the normalized DMO simulations all have $\sigma \simeq 140 \hspace{0.1cm} \kms$ at $r = 400$ pc, the FIRE runs have $\sigma \simeq 350-550 \, \kms$ at the same radius. While this scatter is interesting to note, giving a precise and detailed answer as to why it occurs while require more analysis and is the topic for future analysis. For this present paper, we note that it exists, and that is has a significant impact on the magnitude of the J-factor signal, so that the total magnitude of the J-factor signal varies quite signficantly from halo to halo.

Figure \ref{fig:ratios} shows FIRE to DMO ratios for the normalized density profile (left) and velocity dispersion profile (right) of each halo pair. On the left we see that galaxy formation has generally made the halos less dense at large radii, corresponding to steeper (contracted) density profile.  The effect of galaxy formation on the inner density is quite varied, with some systems (e.g. \texttt{Romulus} and \texttt{Romeo}) displaying higher central densities in the FIRE runs, while others (e.g. \texttt{Thelma} and \texttt{m12m}) have lower densities. There is no clear trend with stellar mass or virial mass associated with these differences.  \texttt{Thelma} and \texttt{Romeo}, for example, have very similar galaxy masses and virial masses but galaxy formation seems to have had an opposite effect on their relative density profiles.This is likely an artifact of some of the important baryonic differences between halos at late times, which are studied in other papers: some have late-occurring minor mergers or strong stellar bars \citep[which tend to push DM outwards and lower central densities; e.g.][]{Sanderson2017,Debattista2019},  others have strong torques or early multiple-mergers which produce inflows and dense bulges and more compact disks \citep[e.g.][]{SGK18,Ma17}. For the purposes of this paper, it is noted that it impacts the calculated J-factor and affects variance from halo to halo. 

The right panel of Figure \ref{fig:ratios} shows again that the effect of galaxy formation on the dark matter velocity dispersion is systematic.  In every case the FIRE runs are hotter, with $\sim 3-4$ times higher velocity dispersion than their DMO counterparts at $r = 400$ pc.  In Appendix B we show that these halos become baryon dominated within 3-8 kpc from the center (Figure \ref{fig:ratios}.  As discussed next, this enhancement in central velocity dispersion has a systematic effect on  the dark-matter annihilation J-factors for velocity-dependent cross sections.

\section{Astrophysical J-Factors}

\subsection{Definitions}

If dark matter particle of mass $m_\chi$ is its own antiparticle with an annihilation cross section $\sigma_A$, the resulting differential particle flux produced by annihilation in a dark matter halo 
can be written as the integral along a line of sight $\ell$ from the observer (located at the solar location in our case) in a direction $\vec{\theta}$ in the plane of the sky over pairs of dark matter particles with velocities $\vec{v}_1$ and $\vec{v}_2$:
\begin{equation}
  \frac{d^2\Phi}{dE d\Omega} =
  \frac{1}{4\pi} \frac{d N}{d E} \int \dd \ell ~\dd^3v_1 \, \dd^3 v_2 
   \frac{f(\vec{r}, \vec{v}_1)}{m_\chi}
  \frac{f(\vec{r}, \vec{v}_2)}{m_\chi}
  \, \frac{(\sigma_A v_{\rm rel}) }{2} \ .
\end{equation}
Here, $\vec{r} = \vec{r}(\ell,\vec{\theta})$ is the 3D position, which depends on the distance along the line of sight $\ell$ and sky location $\vec{\theta}$.  The dark matter velocity distribution $f(\vec{r}, \vec{v})$  is normalized such that $\int d^3v f(\vec{r}, \vec{v}) = \rho (\vec{r})$, where $\rho$ is the dark matter density at that location.  The symbol $v_{\rm rel}=|\vec{v}_1 - \vec{v}_2|$ represents the relative velocity between pairs of dark matter particles.  The quantity $m_\chi$ is the dark matter particle mass and $d N / d E$ is the particle energy spectrum ultimately produced by a single annihilation.  

Following \citet{Boddy18}, we parameterize the velocity-dependence of the dark matter annihilation cross section as 
\begin{equation}
    \sigma_A v_{\rm rel} = [\sigma v]_0 \, Q(v_{\rm rel}),
    \label{eq:sigv}
\end{equation}
where $[\sigma v]_0$ is the overall amplitude and the function $Q(v)$ parameterizes the velocity dependence. For $s$-wave, $p$-wave, and $d$-wave annihilation, $Q(v) = 1$, $(v/c)^2$, and $(v/c)^4$, respectively.  We can then  rewrite the differential particle flux as
\begin{equation}
  \frac{d^2 \Phi}{d E \dd \Omega} =
  \frac{(\sigma_A v)_0}{8\pi m_X^2} \frac{d N}{d E_\gamma} \left[ \frac{d J_Q}{d  \Omega} \right]  \, .
\label{eq:Jfactor}
\end{equation}
Here, the term in brackets absorbs all of the astrophysics inputs and defines the astrophysical "J-factor" 
\begin{equation}
  \frac{dJ_Q}{d \Omega} (\vec{\theta}) = 
  \int \dd \ell \int \dd^3 v_1 {f(\vec{r}, \vec{v}_1)}
  \int \dd^3 v_2 {f(\vec{r}, \vec{v}_2)}\, Q( v_{\rm rel}) \, .
  \label{eq:Jfactor_def}
\end{equation}
In principle, the $\ell$ integral above sums pairs along the line-of-sight from the observer ($\ell = 0$) to infinity. In practice, we are focusing on J-factors arising from an individual ``Milky Way" halo, and truncate our integrals at the halo's edge (see below).

It is often useful to quote the cumulative J-factor within a circular patch of sky of angular radius $\psi$ centered on the Galactic Center.  In this case, the patch defined by $\psi$ subtends a solid angle $\Omega_\psi = 4 \pi \sin^2({\psi}/2)$ and we have:
\begin{equation}
  J_Q (< \psi)   =  \int_0^{\Omega_\psi} \frac{d J_Q}{d  \Omega} (\vec{\theta}) \,  \dd \Omega \, .
  \label{eq:Jf}
\end{equation}

\begin{figure*}
	\includegraphics[width=\columnwidth,trim = 130 0 50 90]{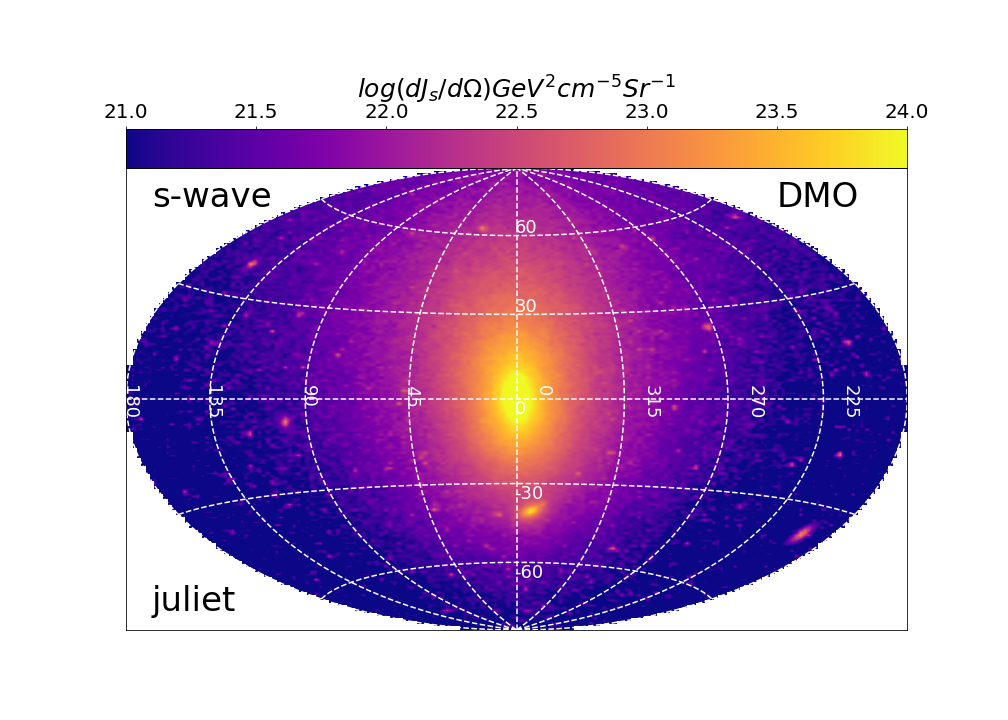}
		\includegraphics[width=\columnwidth, trim = 50 0 130 90]{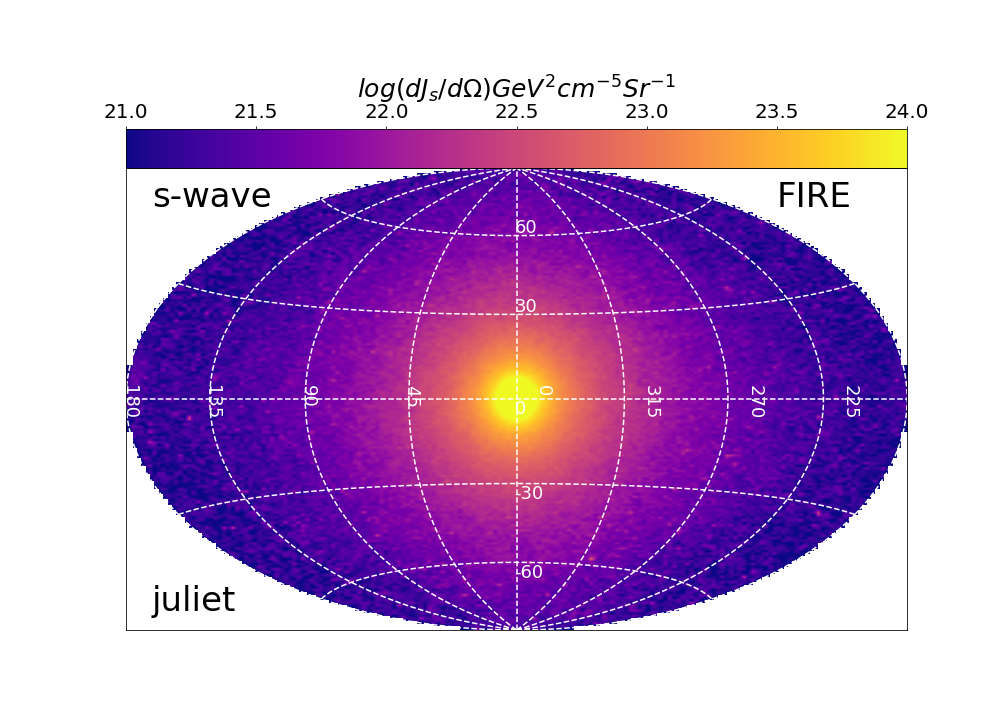} \\
	\includegraphics[width=\columnwidth,trim = 130 0 50 90]{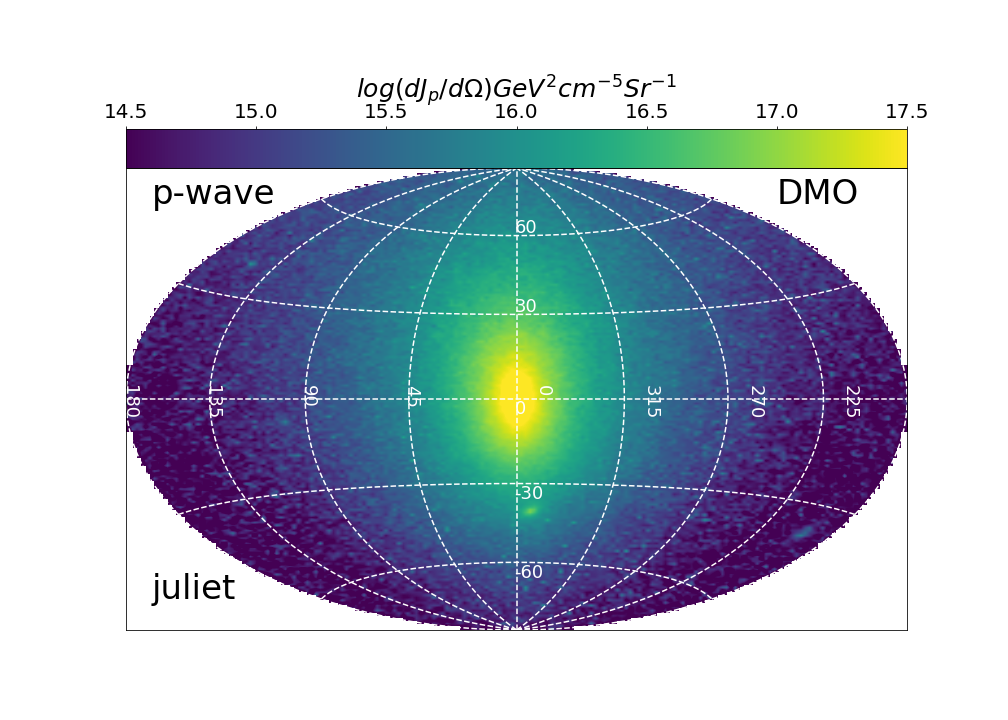}
		\includegraphics[width=\columnwidth, trim = 50 0 130 90]{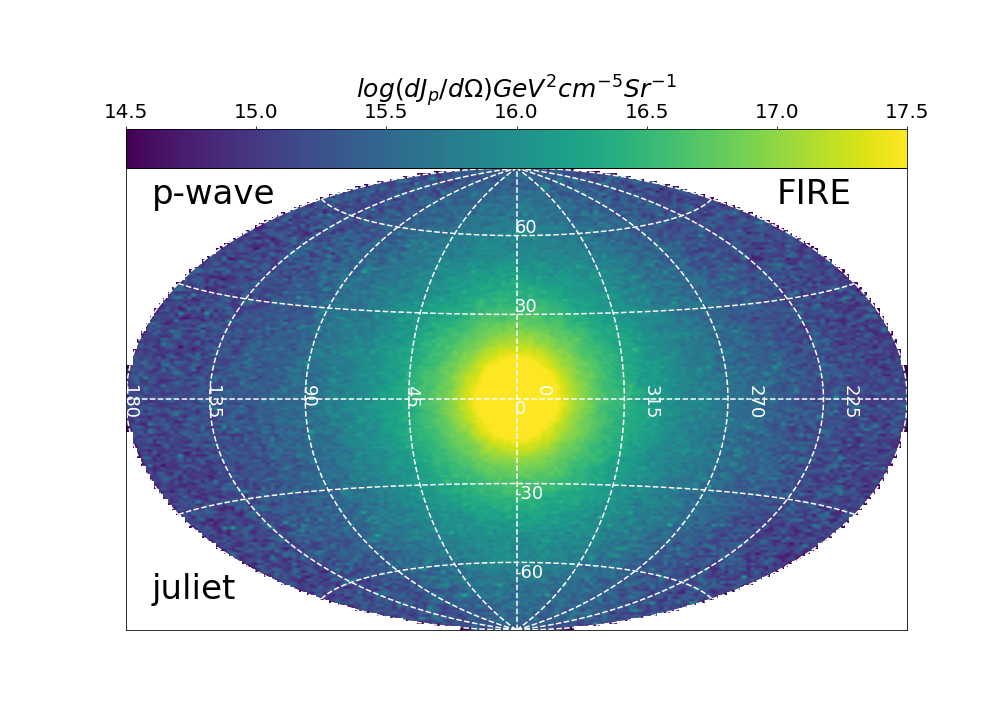} \\
	\includegraphics[width=\columnwidth,trim = 130 0 50 90]{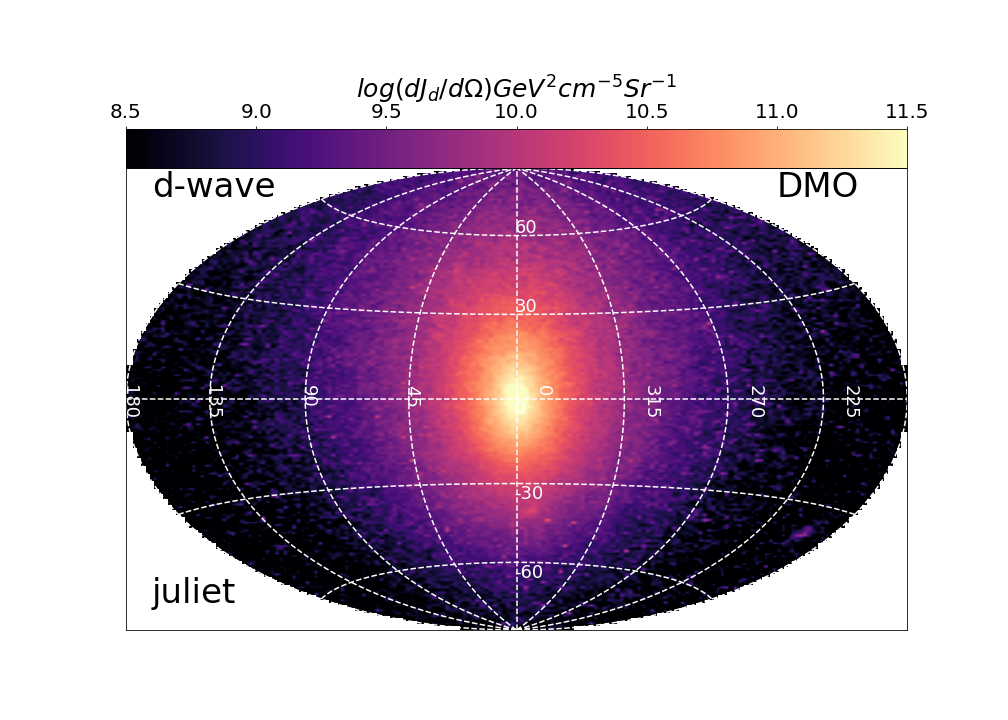}
		\includegraphics[width=\columnwidth, trim = 50 0 130 90]{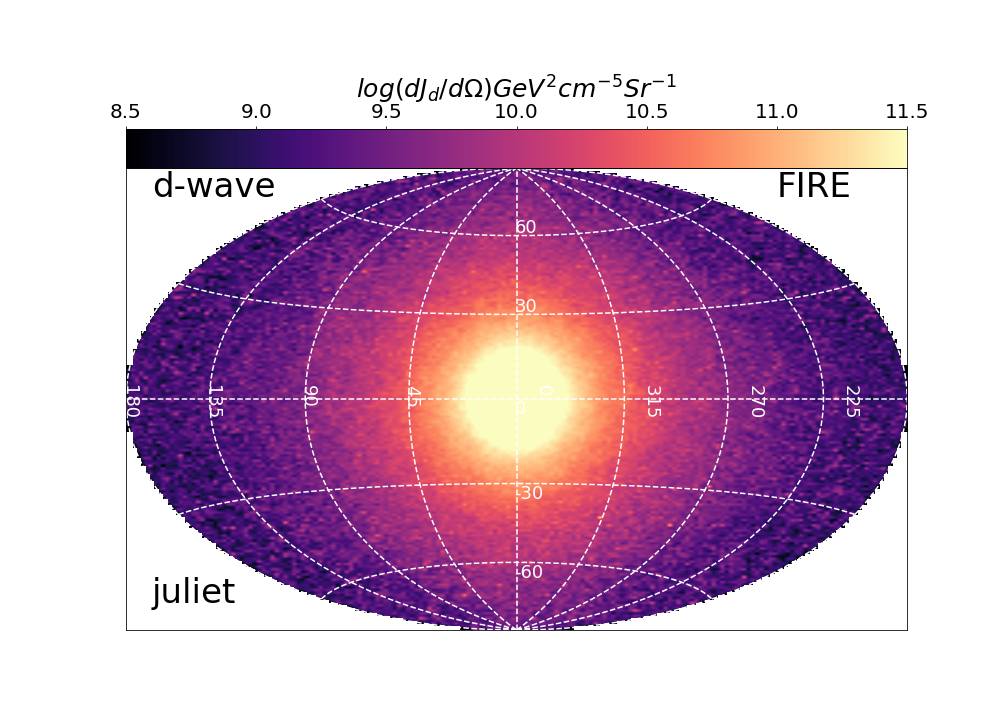}
    \caption{All-sky Hammer projections of J-factors ($d J/d \Omega$)  for \texttt{JulietDMO} (left)  and \texttt{Juliet} (right) in Galactic coordinates as viewed from mock solar locations 8.3 kpc from the halo centers.  Maps utilize bins of roughly 1.3 square degrees on the sky. From top to bottom, the rows assume s-wave, p-wave, and d-wave annihilation.  The color map in each pair of panels is fixed for each type of assumed annihilation and is logarithmic in $d J/d \Omega$, as indicated by the bar along the top of each image.  Note that FIRE runs (right) produce systematically rounder J-factor maps on the sky.  The p-wave and d-wave maps are significantly brighter and more extended for the FIRE runs as well, owing to the effects of galaxy formation in enhancing dark matter velocity dispersion in the center of each halo.}
    \label{fig:map_juliet}
\end{figure*}

\begin{figure*}
	\includegraphics[width=\columnwidth,trim = 130 0 50 90]{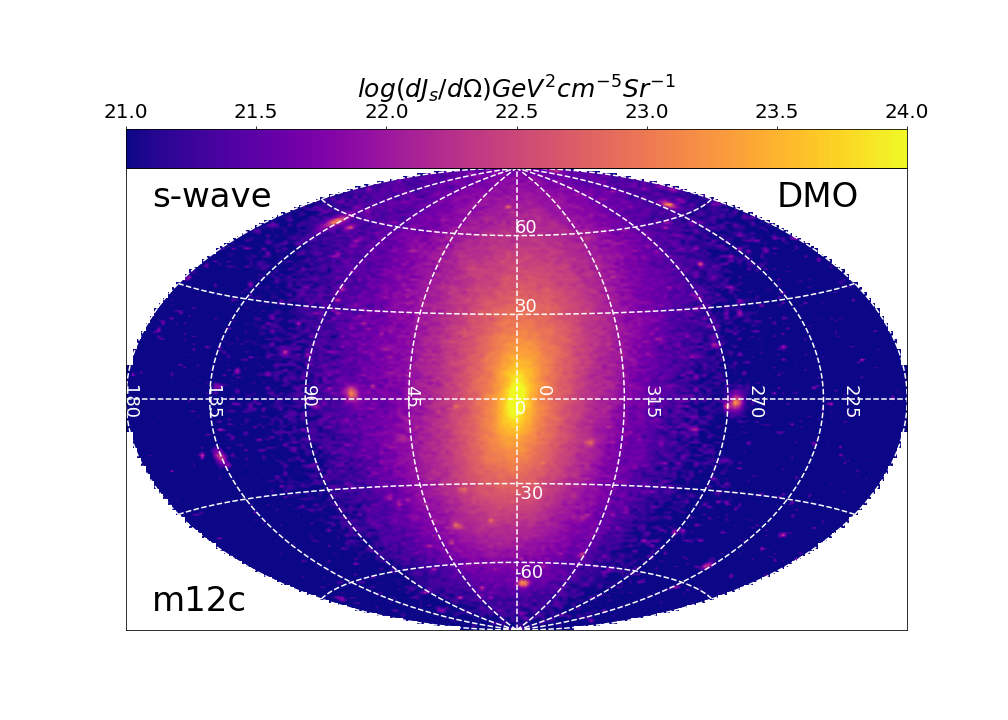}
		\includegraphics[width=\columnwidth, trim = 50 0 130 90]{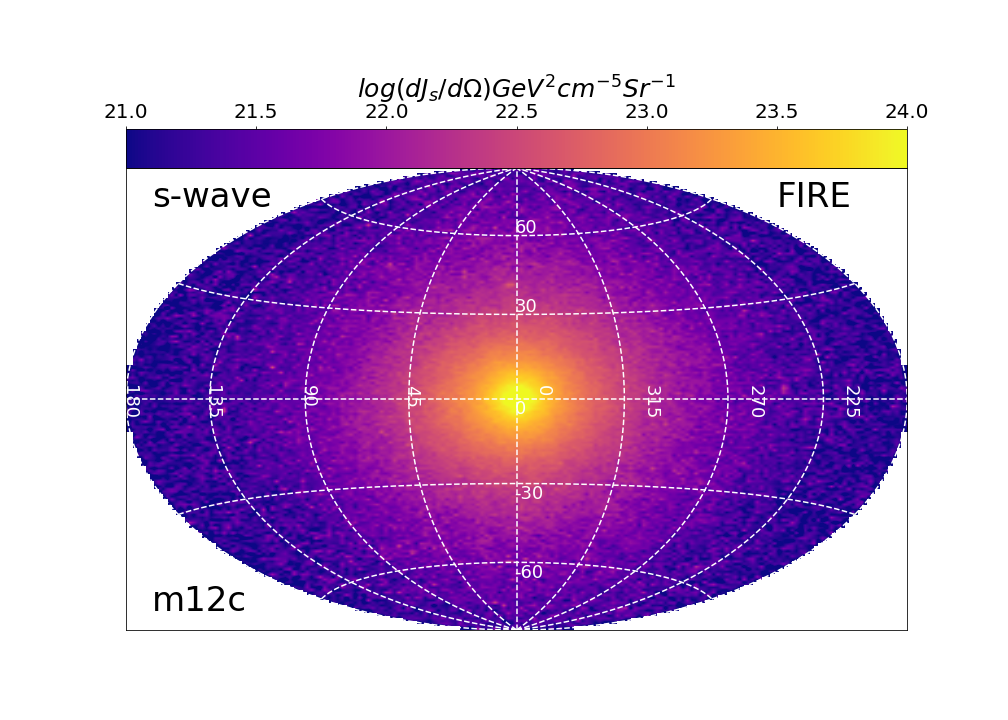} \\
	\includegraphics[width=\columnwidth,trim = 130 0 50 90]{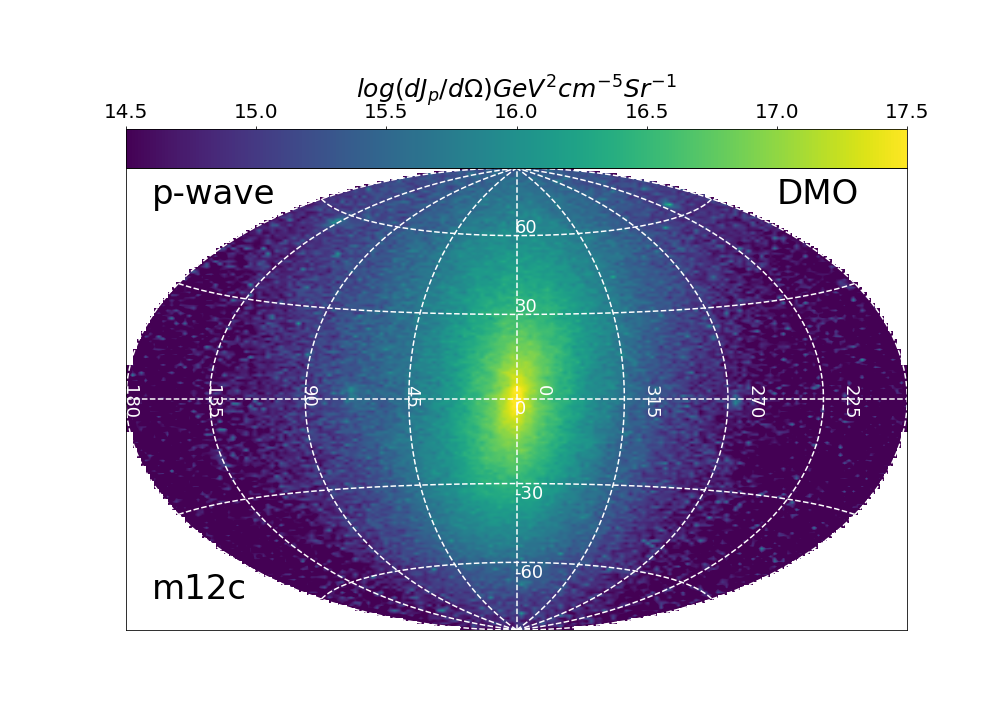}
		\includegraphics[width=\columnwidth, trim = 50 0 130 90]{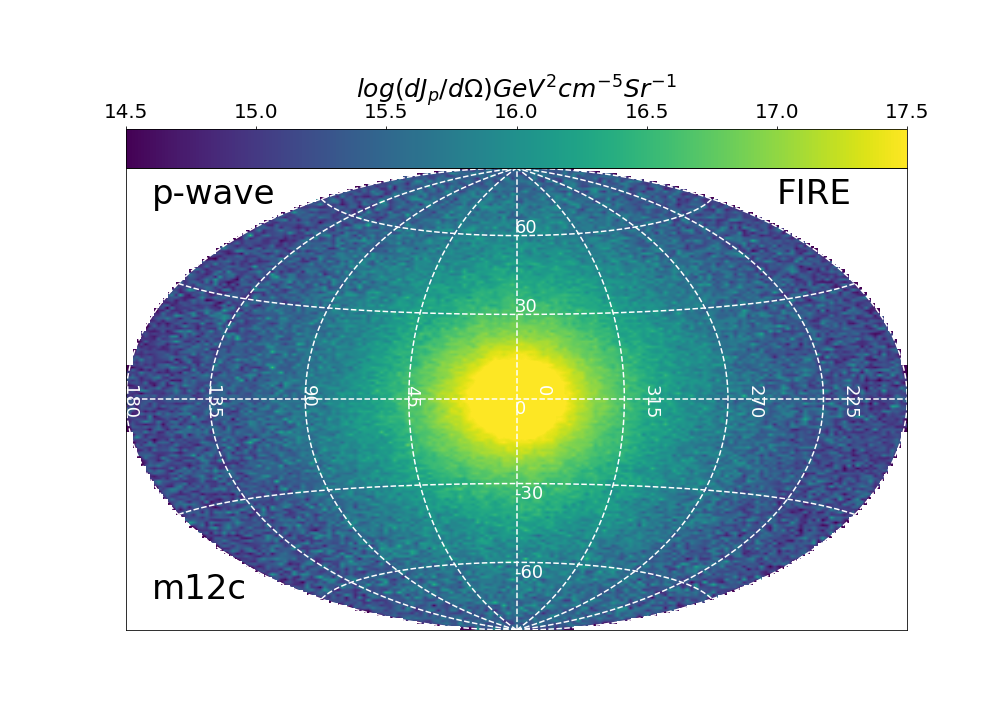} \\
	\includegraphics[width=\columnwidth,trim = 130 0 50 90]{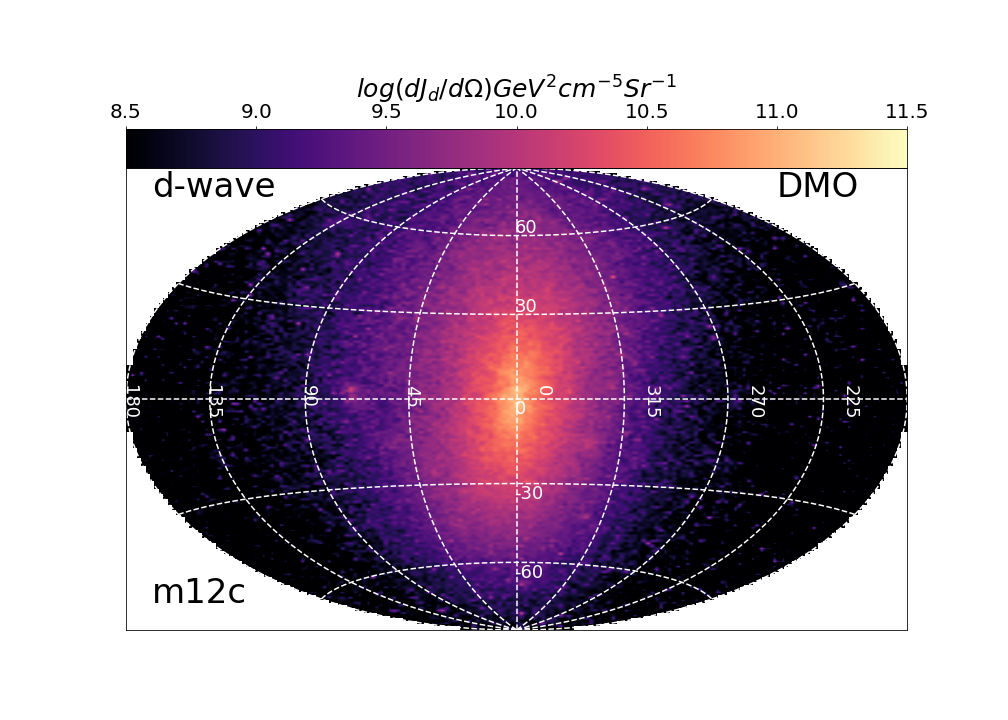}
		\includegraphics[width=\columnwidth, trim = 50 0 130 90]{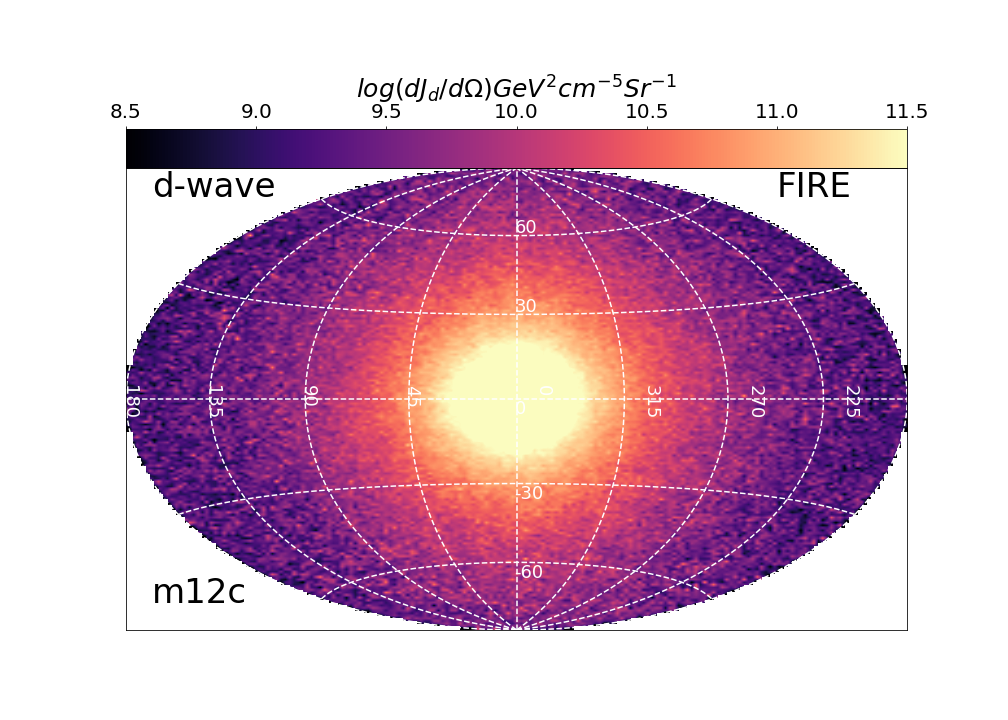}
    \caption{Same as figure 3, but for \texttt{M12cDMO} (left)  and \texttt{M12c} (right).} 
    \label{fig:map_m12c}
\end{figure*}

\subsection{Approach}
\label{sec:approach}

In what follows we aim to determine the astrophysical J-factors for each of our simulated halos for s-wave, p-wave, and d-wave annihilation.  In doing so we approximate the dark matter distribution $f(\vec{r}, \vec{v})$ as a separable function:
\begin{equation}
    f(\vec{r}, \vec{v}_1) = \rho(\vec{r}) \, g(\vec{v}(\vec{r})),
\end{equation}
with the dark matter density $\rho$ estimated using direct particle counts in the simulation. In this estimate, we use a cubic spline smoothing kernel \citep{Monaghan92} with smoothing length set to contain the mass of the nearest $32$ neighbors \citep[as described in][]{Hopkins15}.

For standard $s$-wave annihilation we have $Q(v)=1$ and the effective J-factor (Eq. \ref{eq:Jfactor_def}) reduces to a simple integral over the density squared:
\begin{eqnarray}
\frac{d J_{s}}{d \Omega} (\vec{\theta}) & = &
  \int \dd \ell \, \rho^2(\vec{r}) \int \dd^3 v_1 \, g_r(v_1)
  \int \dd^3 v_2 \, g_r(v_2)   \nonumber \\
   & = &  \int \dd \ell \, \rho^2[\ell(\vec{r})].
\end{eqnarray}
For $p$-wave annihilation, $Q(v) = (v/c)^2$, and Eq. \ref{eq:Jfactor_def} becomes
\begin{eqnarray}
\frac{d J_{p}}{d \Omega} (\vec{\theta}) & = &
  \int \dd \ell \, \rho^2(\vec{r}) \int \dd^3 v_1 \, g_r(v_1)
  \int \dd^3 v_2 \, g_r(v_2) \, \frac{|\vec{v}_1 - \vec{v_2}|^2}{c^2}  \nonumber \\
   & = & \frac{1}{c^2} \int \dd \ell \, \rho^2[\ell(\vec{r})] \, \mu_2(\ell(\vec{r})).
   \label{eqn:dp}
\end{eqnarray}
In the second line we have used $\mu_2$ to represent the second moment of the relative velocity at position $\vec{r}$.   
For $d$-wave annihilation, $Q(v) = (v/c)^4$, which implies
\begin{eqnarray}
\frac{d J_{d}}{d \Omega} (\vec{\theta}) & = &
  \int \dd \ell \, \rho^2(\vec{r}) \int \dd^3 v_1 \, g_r(v_1)
  \int \dd^3 v_2 \, g_r(v_2) \, \frac{|\vec{v}_1 - \vec{v_2}|^4}{c^4}  \nonumber \\
   & = & \frac{1}{c^4} \int \dd \ell \, \rho^2[\ell(\vec{r})] \, \mu_4(\ell(\vec{r})).
   \label{eqn:dd}
\end{eqnarray}
Here $\mu_4$ is the fourth moment of the relative velocity at position $\vec{r}$. We measure both $\mu_2$ and $\mu_4$ at each particle position using the nearest 32 dark matter particles.~ \footnote{For a perfectly spherically symmetric Maxwellian distribution, we expect the cross terms to vanish such that $\langle (\vec{v_1} - \vec{v_2})^2 \rangle  = 2 \sigma_v^2$ and $\langle (\vec{v_1} - \vec{v_2})^4 \rangle = 48 \sigma_v^4$ / 9, where $\sigma_v$ is the local velocity dispersion. We show in {Section  \ref{sec:maxwell} that direct measurement gives slightly lower estimates than would be expected from the simplified Maxwellian expectation.  We then construct all-sky maps of the relevant J-factors using appropriately-weighted and smoothed projections from mock observer locations (see below).}} 

\subsection{Geometric Setup}

For each halo in our sample, we calculate J-factors as defined in Equation \ref{eq:Jfactor_def}, integrating from a mock Solar location (setting $\ell = 0$) to the edge of the halo, which we define as a sphere of radius $r = 300$ kpc from the center of each halo in every case.  While the virial radii \citep{ByranNorman1998} of our halos range from $300 - 335$ kpc, we fix $300$ kpc as the halo boundary for consistency.  Since most of the J-factor signal comes from the inner halo, changing the outer radius by $10\%$ has no noticeable affect on our results. 

For the DMO runs, we assume that the Galactic Center corresponds to the halo center and fix the observer location to be at a distance 8.3 kpc from the halo center along the x-axis of the simulation.  For FIRE runs, we position the observer in the galaxy disk plane at a radius of 8.3 kpc from the halo center.  We define the disk plane to be perpendicular to the angular momentum vector of all the stars within 20 kpc of the central galaxy. 


\begin{figure*}
		\includegraphics[height =0.6\columnwidth, trim = 0 0 0 0 ]{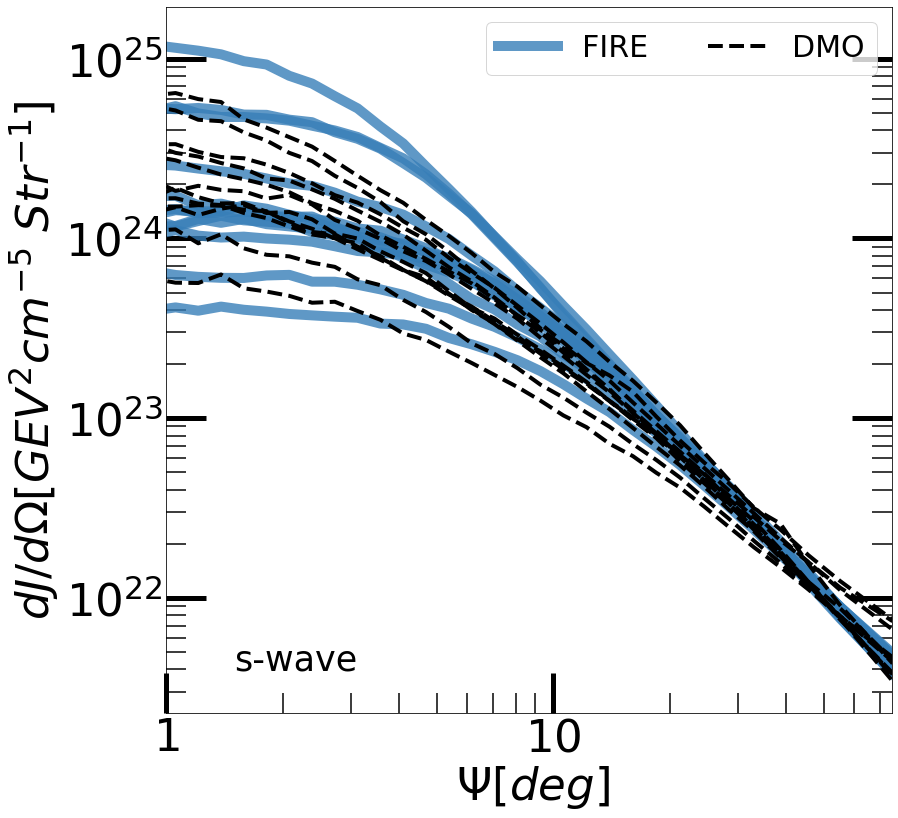} \hspace{.1in}
				\includegraphics[height =0.6\columnwidth, trim = 0 0 0 0 ]{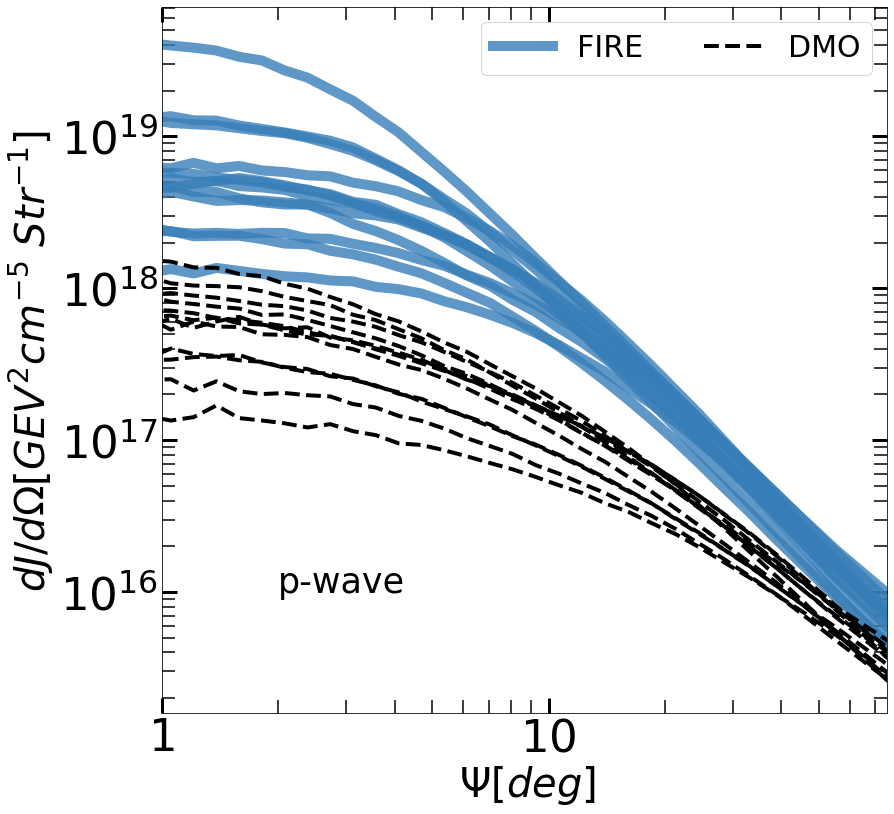}\hspace{.1in}
						\includegraphics[height =0.6\columnwidth, trim = 0 0 0 0 ]{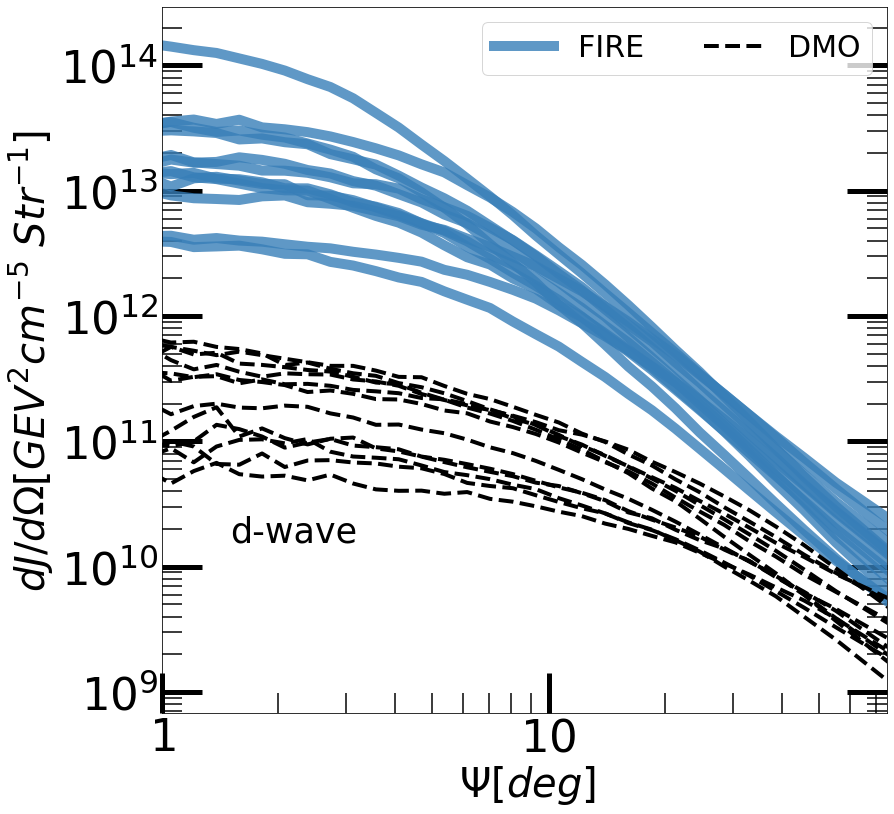}
    \caption{Differential J-factor profiles (Equation \ref{eq:Jfactor_def}) as a function of angle $\psi$ from the Galactic center for s-wave, p-wave, and d-wave annihilation (from left to right).  The FIRE simulation halos are shown as solid blue lines and their DMO counterparts are shown as dashed black lines. For the s-wave case (left) we see that the central J-factor values are similar for FIRE and DMO cases, though the FIRE profiles are flatter at small angle, and more extended on the sky.  The FIRE p-wave profiles (middle) are a noticeably amplified compared to the DMO cases.  Their shapes are also significantly different -- with a flatter inner profile and sharper fall-off at angles beyond 10 degrees.  The d-wave case (right panel) demonstrates the starkest difference between FIRE and DMO runs.  }
    \label{fig:djdo}
\end{figure*}

\begin{figure*}
			\includegraphics[height =0.6\columnwidth,trim = 0 0 0 0]{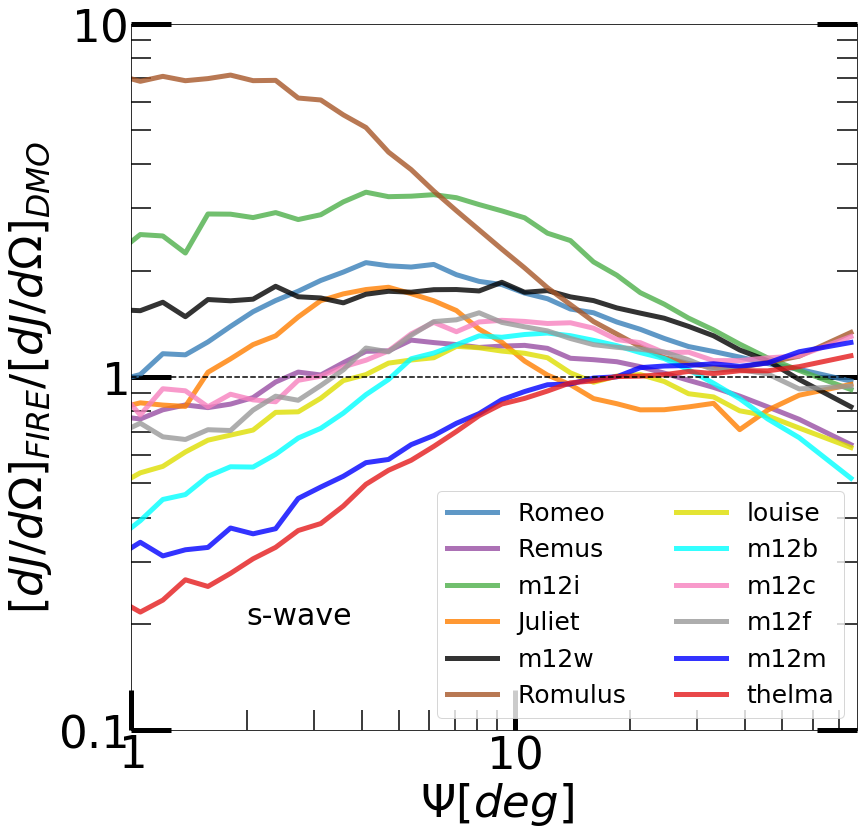} \hspace{.1in}
		\includegraphics[height =0.6\columnwidth, trim = 0 0 0 0]{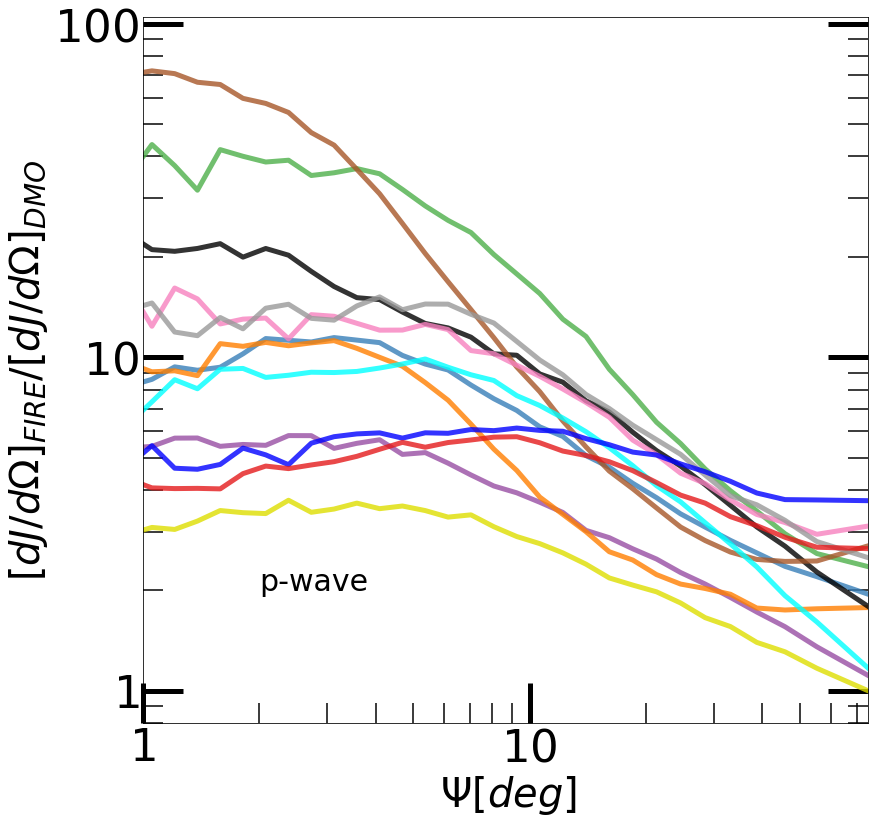} \hspace{.1in}
			\includegraphics[height =0.6\columnwidth, trim = 0 0 0 0]{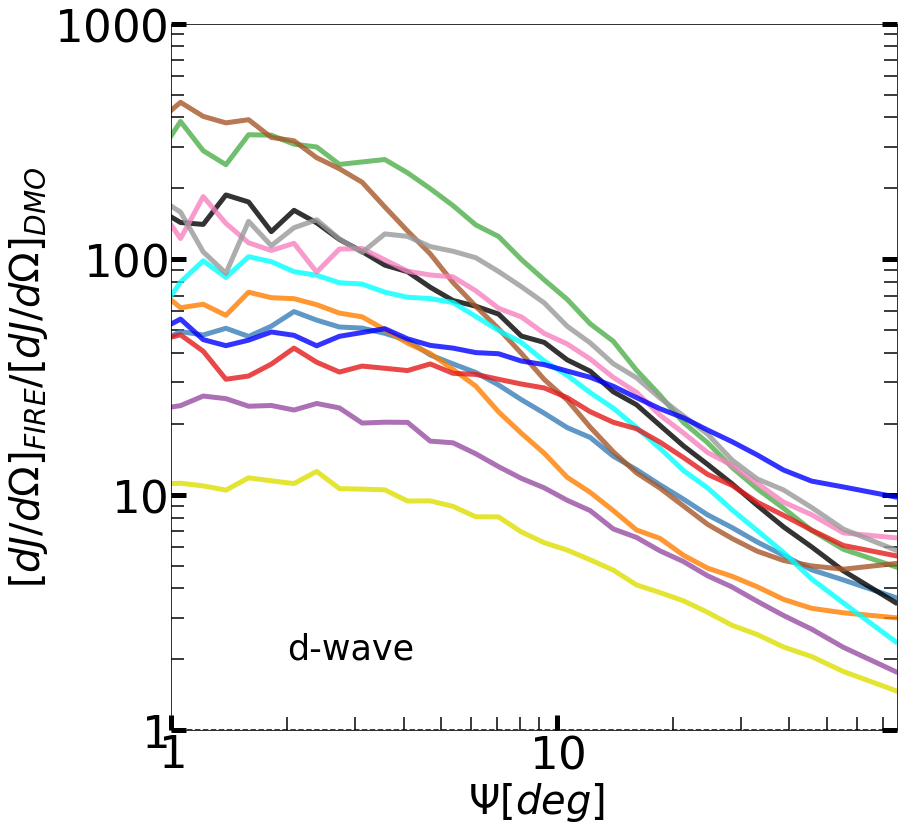}
    \caption{Ratio of FIRE to DMO J-factors for each halo as a function of  angle $\psi$ from the Galactic center for s-wave, p-wave, and d-wave annihilation (from left to right).  Each halo pair has a unique color, as indicated.  For the case of s-wave (left), the ratio is of order unity, ranging from a factor of $\sim 3$ higher to a factor of $\sim 0.3$ lower at small angles.  For p-wave, the FIRE runs have J-factors as much as $\sim 30$ times higher, though the amplification can be as small as a factor of $\sim 3$.  In the case of d-wave, amplification factors as large as $\sim 100-400$ at small angle are seen. }
    \label{fig:djratio}
\end{figure*}

\begin{figure*}
    \centering
    \includegraphics[width=\columnwidth]{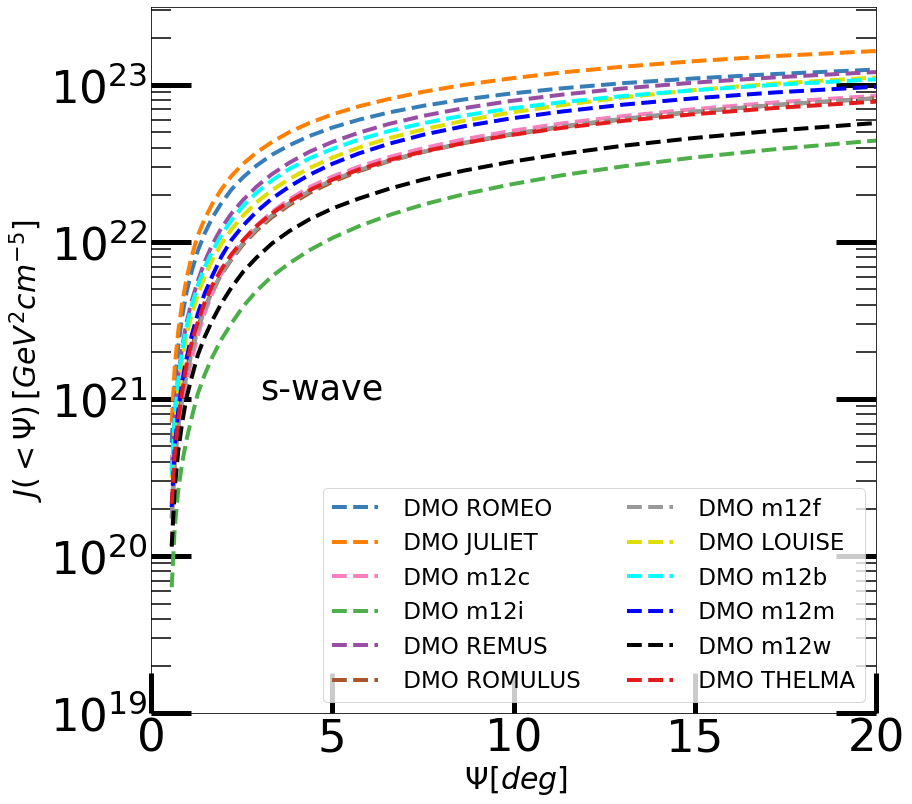} 
	\includegraphics[width=\columnwidth]{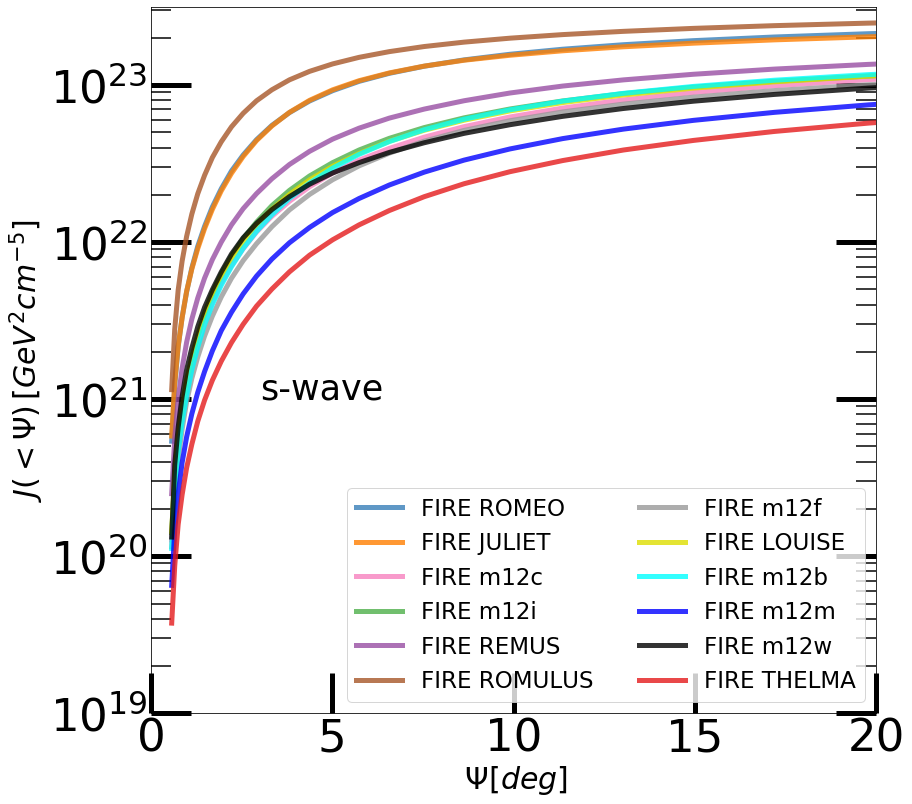} 	\\
	\includegraphics[width=\columnwidth]{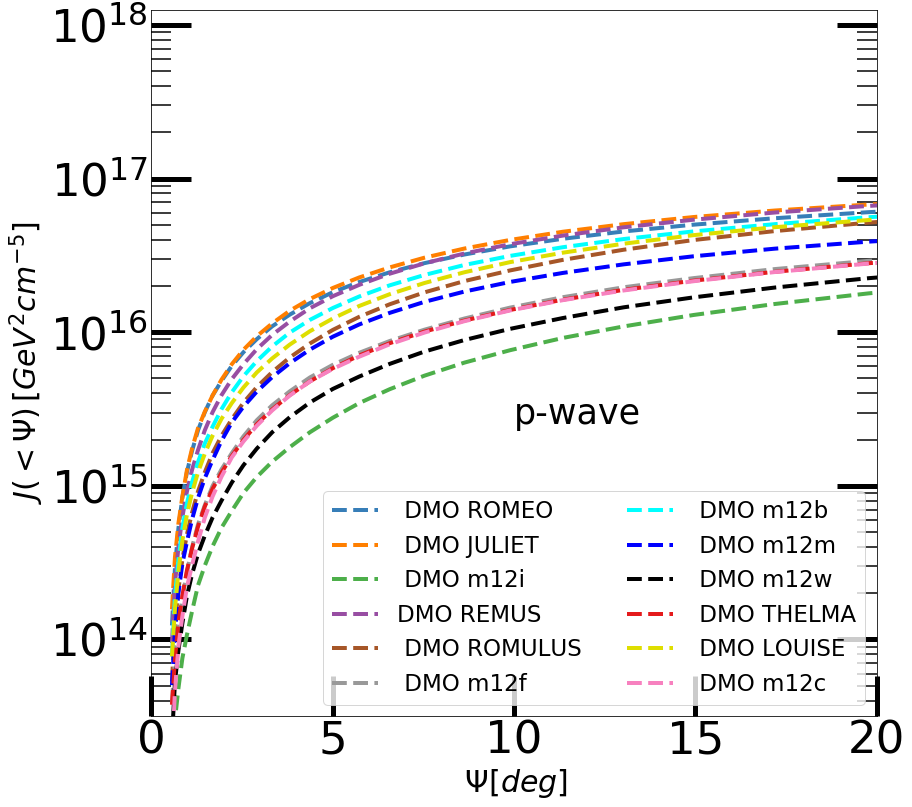}
	\includegraphics[width=\columnwidth]{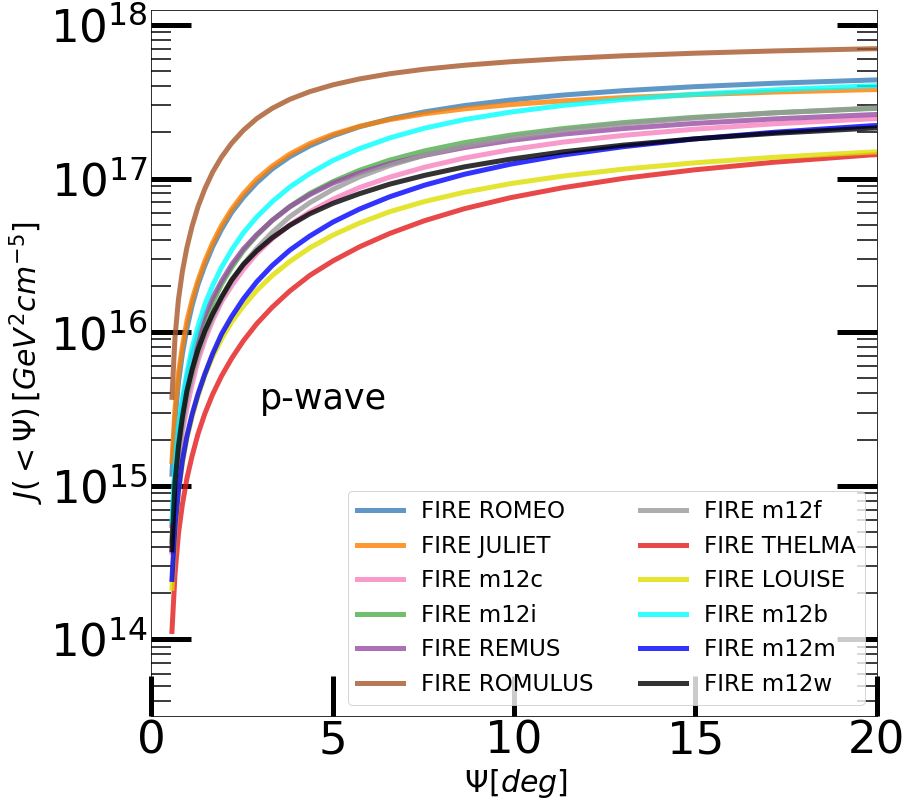} 	 \\
	\includegraphics[width=\columnwidth]{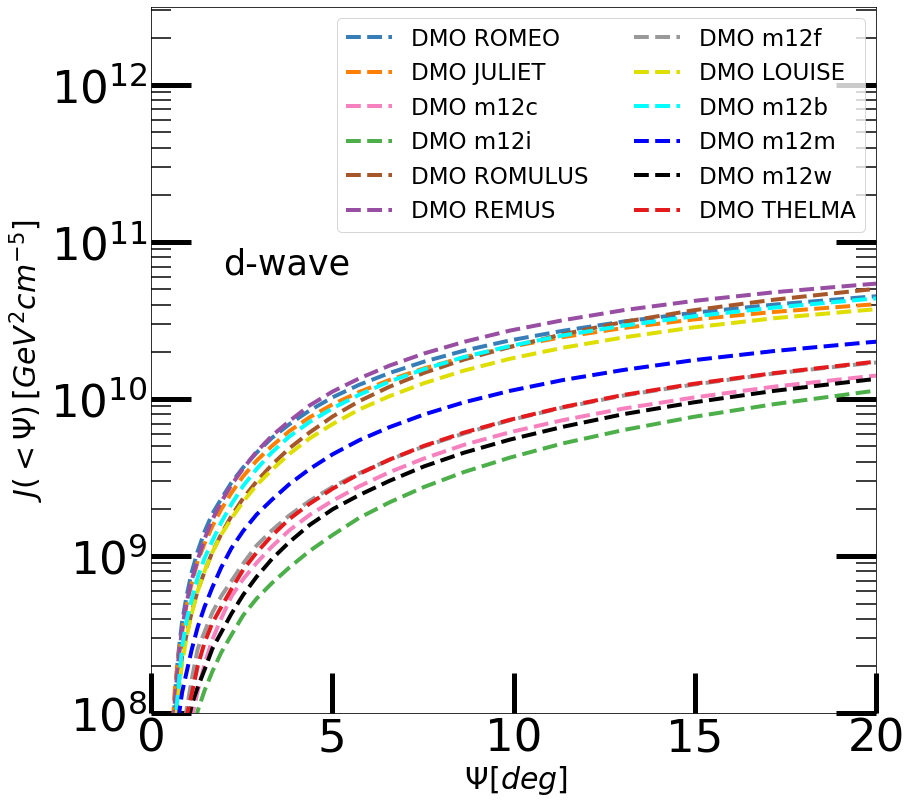}
		\includegraphics[width=\columnwidth]{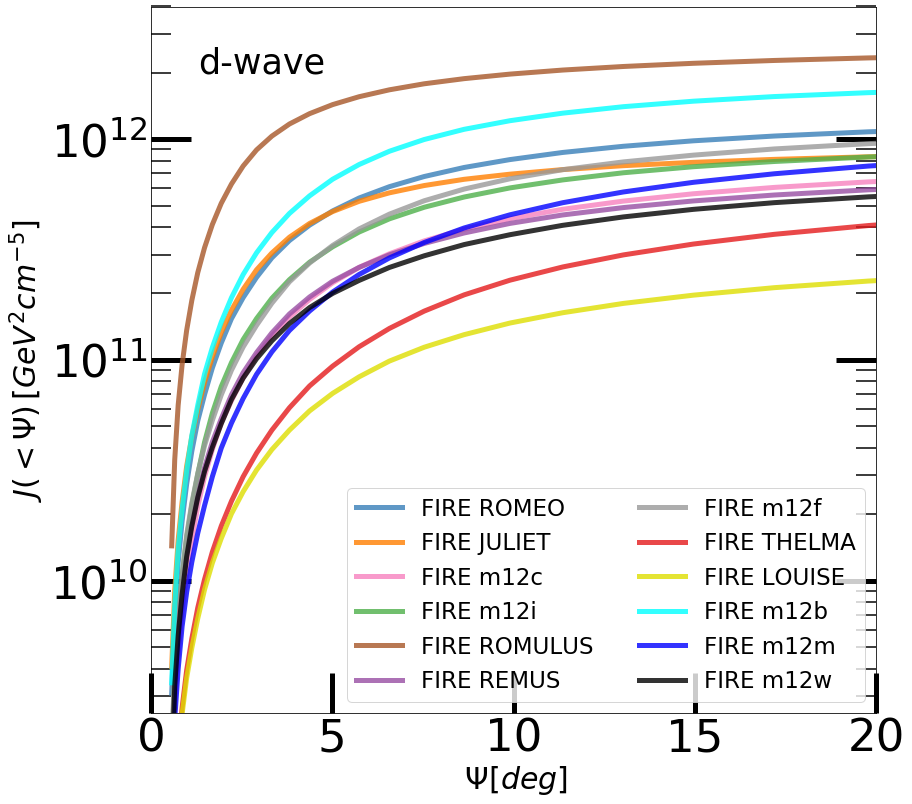} 	 \\
    \caption{  Cumulative J-factor within annular angle $\psi$ from the Galactic Center (Eq. \ref{eq:Jf}) for DMO (left) and FIRE (right) for s-wave (top), p-wave (middle), and d-wave (bottom) annihilation. For the s-wave case, FIRE runs have J-factors of the same order of magnitude as the DMO runs, while for p-wave and d-wave the J-factors are considerably higher.} 
   \label{fig:Jtot}
\end{figure*}

\begin{figure*}
    \centering
    \includegraphics[width=\columnwidth]{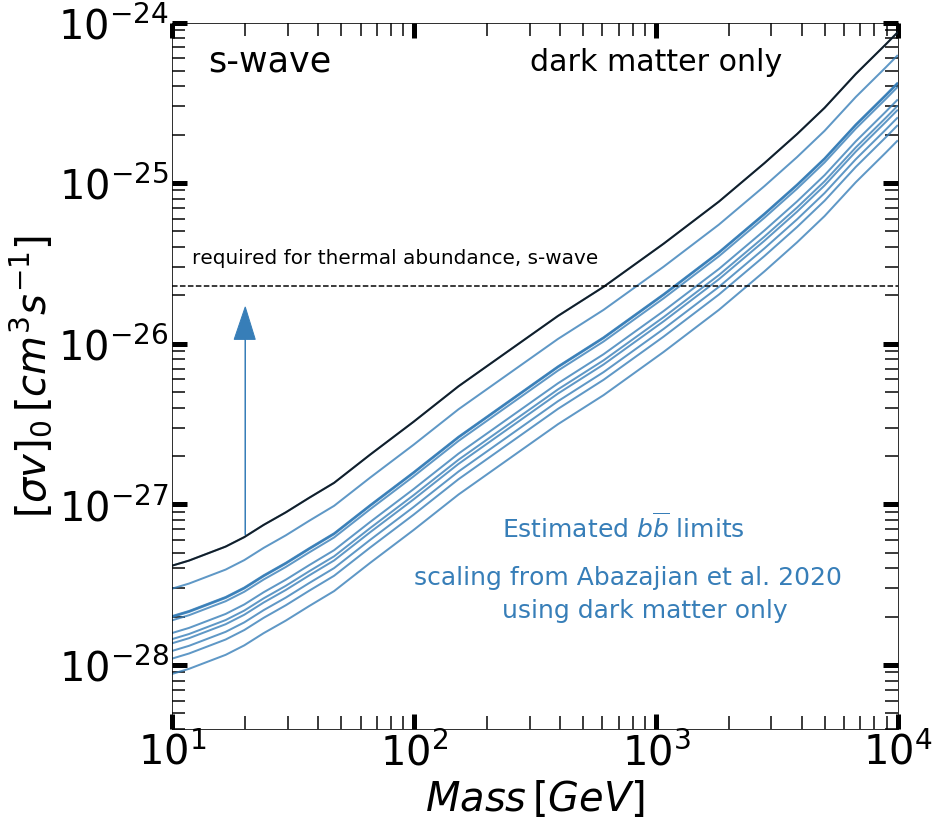} 
	\includegraphics[width=\columnwidth]{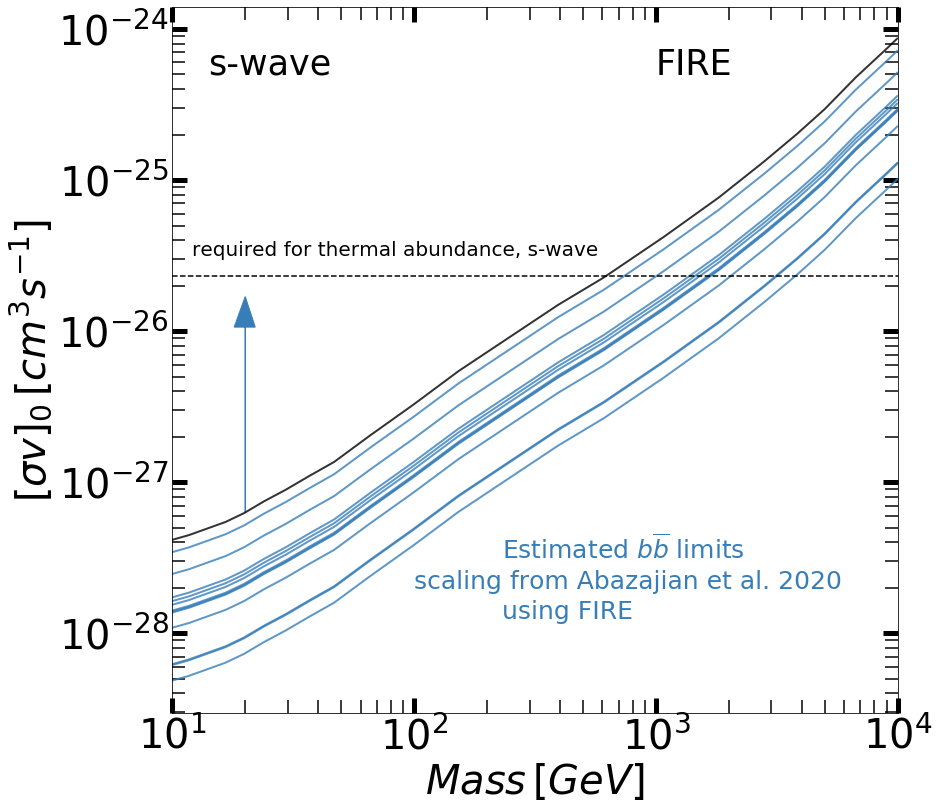} 	\\
	\includegraphics[width=\columnwidth]{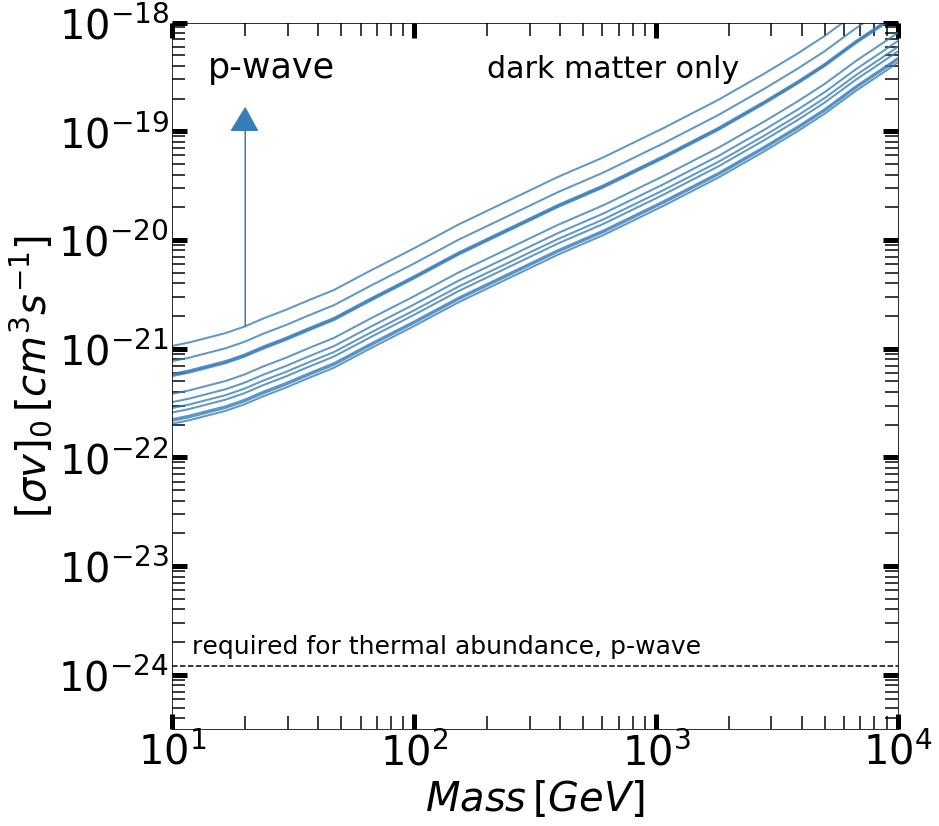}
	\includegraphics[width=\columnwidth]{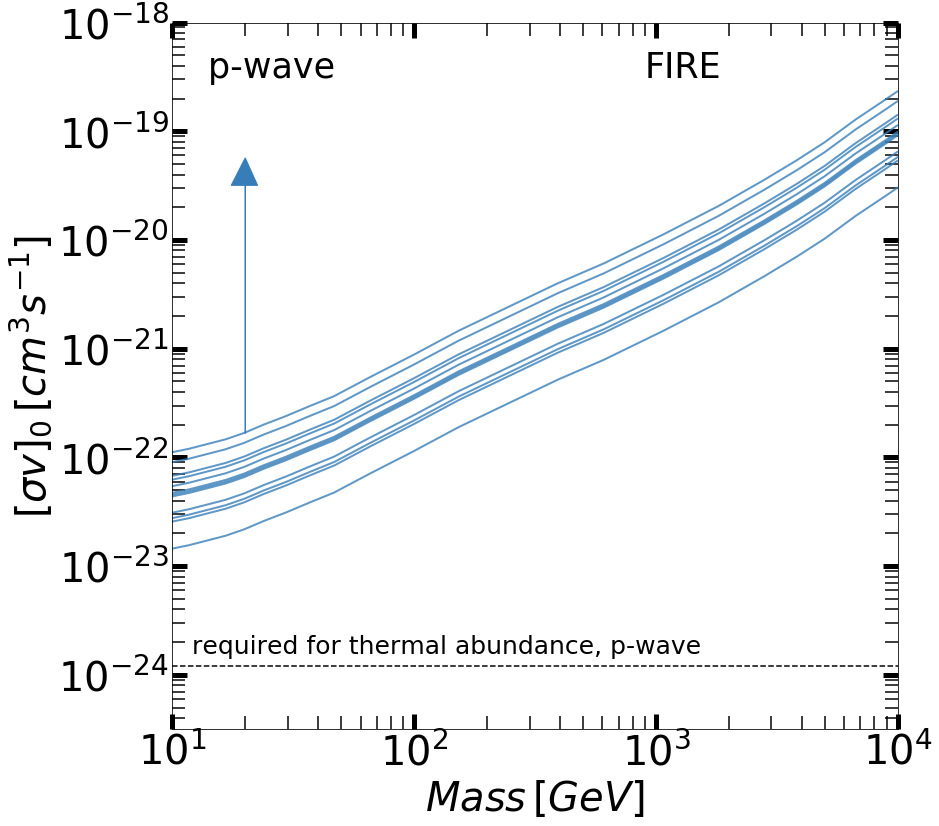} 	 \\
	\includegraphics[width=\columnwidth]{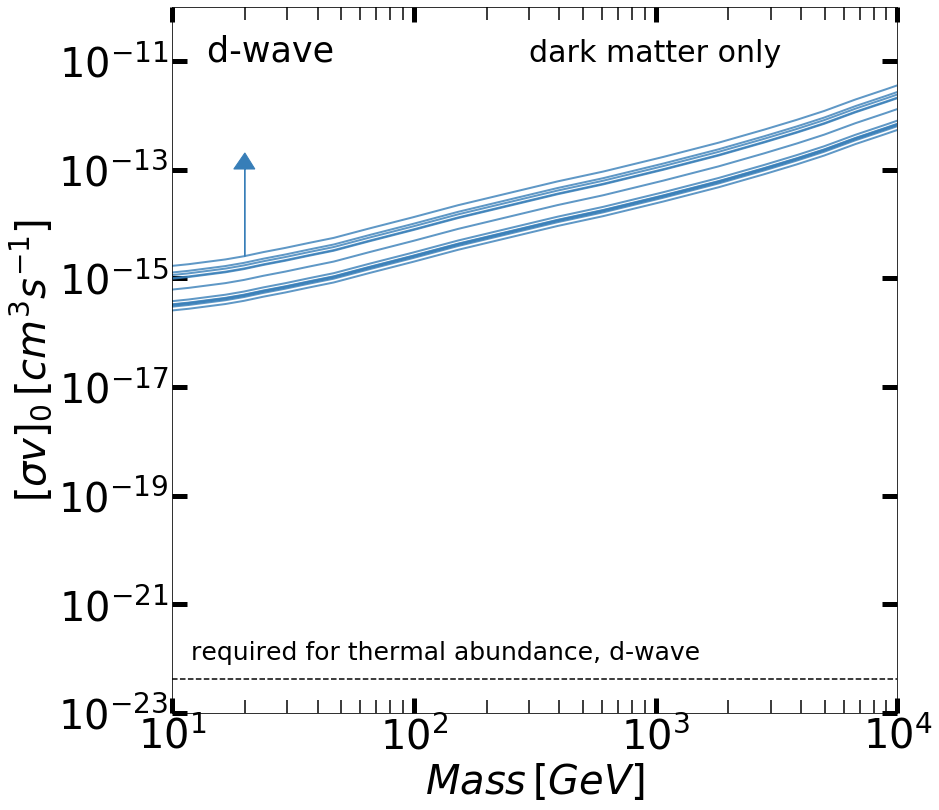}
		\includegraphics[width=\columnwidth]{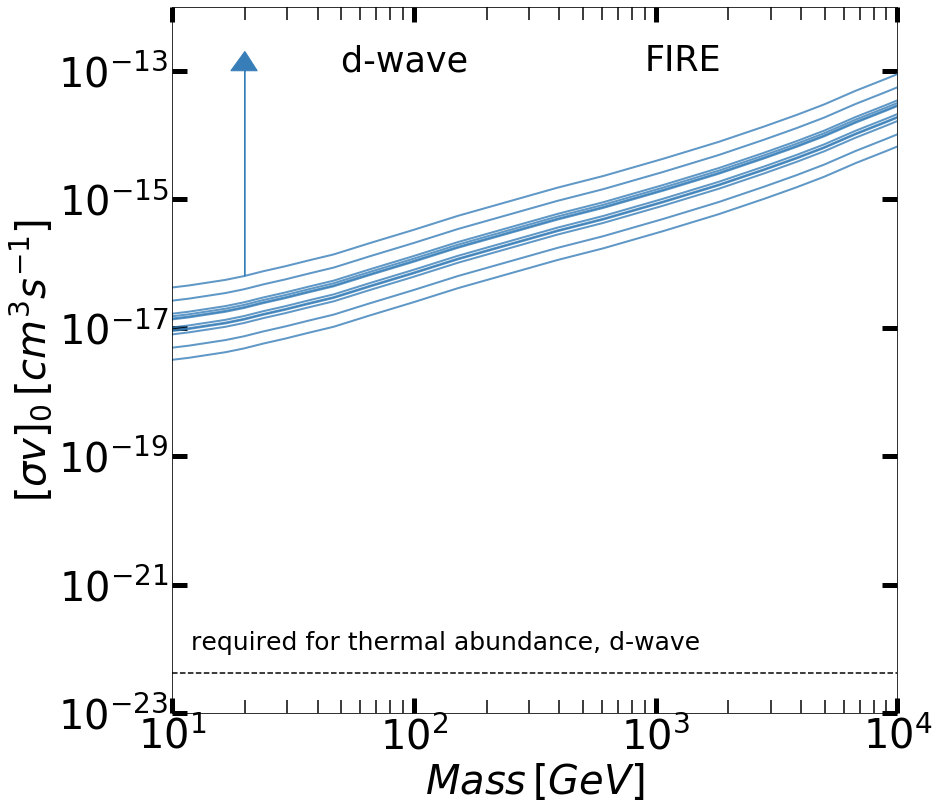} 	 \\
    \caption{Schematic illustration of how the cross section versus particle mass constraints from \citet{Abazajian20} (black solid lines, upper panels) would shift for s-wave (top), p-wave (middle), and d-wave (bottom) annihilation based on the relative J-factors (blue) from each of our DMO (left) and FIRE runs (right).  The region above the lines is ruled out.  The dotted lines show the $[\sigma v]_0$ required for thermal dark matter to match the observed abundance.  FIRE results suggest that current constraints for p-wave are much closer to the thermal cross section than would have been expected from DMO halos, potentially within a factor of $\sim 10$ for $10$ GeV WIMPS.} 
    \label{fig:constraints}
\end{figure*}


\section{Results}

Figures \ref{fig:map_juliet} and \ref{fig:map_m12c} illustrate graphically our results for two example halo pairs, \texttt{Juliet} and \texttt{m12c}, respectively.  We show all-sky Hammer projection maps of  $dJ/d\Omega$ for s-wave (top), p-wave (middle), and d-wave (bottom) for the DMO (left) and FIRE (right) run of each halo.  We are viewing the Galactic Center (middle of each image) from mock solar locations as defined in the previous section.  The color bars are mapped to J-factor amplitude as indicated at the top of each image.  Note that every row has the same color mapping, so that the relative difference between DMO and FIRE runs can be seen clearly for each assumed velocity dependence.  The binning in these maps is $1.3$ square degrees.To find the local density of the particles and the nearest neighbors for our velocity calculations, we used Firestudio \citep{FIRESTUDIO}.

The first takeaway from these images is that the FIRE runs are significantly brighter (with amplified J-factors) than the DMO runs for the p-wave and d-wave cases.  This is a direct result of the FIRE halos having enhanced dark matter velocity dispersion compared to the DMO halos (e.g., the right panel of Figure \ref{fig:ratios}). The maps are also more extended from the Galactic Center. 
The s-factor maps are not as different, given the modest differences in central densities for these particular halos (see the left panel of Figure \ref{fig:ratios}), though substructure is significantly reduced in the FIRE runs, as expected from the destructive effects of the central galaxy \citep{garrison2017not,Kelly19}. Note that in all cases, including s-wave, the FIRE maps are {\em rounder} on the sky -- this is a result of galaxy formation tending to sphericalize the dark matter distributions compared to DMO runs in halo centers \citep{Debattista08,bernal2016spherical,Chau19,Kelly19,Shen21,Sameie21}.  The fact that we expect annihilation signals to be even rounder than in the DMO case should in principle make it easier to detect or exclude dark matter annihilation in the face of astrophysical backgrounds, which are expected to track more closely the shape of the Galaxy \citep[e.g.][]{Abazajian20}. We will not focus on quantifying this difference in shape on the sky here because it will be the subject of future work.  We will instead focus on azimuthally-averaged results in what follows.

Figure \ref{fig:djdo} provides a summary of J-factor results for all of our FIRE halos (solid blue) and DMO halos (dashed black). Plotted are $dJ/d\Omega$ profiles (Equation \ref{eq:Jfactor_def}) as a function of angle $\psi$ with respect to the Galactic Center. Results for s-wave, p-wave, and d-wave are shown in separate panels, from left to right. As expected, the FIRE runs are generally amplified compared to DMO runs, especially for the p-wave and d-wave cases.  The shapes of the profiles are also significantly different in character.  While the DMO runs show a trend for J-factor profiles to be more peaked at small angle for s-wave, and to become flatter and more extended on the sky as we progress to p-wave and d-wave, the FIRE profiles are more similar in shape.  In all cases (s-wave, p-wave, and d-wave) the "emission" profile is fairly constant out to $\sim 5-10$ degrees in the FIRE runs, with a steep fall-off towards larger angles beyond that point.  

Figure \ref{fig:djratio} shows, for each halo pair, the ratio of $dJ/d\Omega$ in FIRE to the DMO case as a function of angle from the Galactic Center, $\psi$.  Each halo pair has a unique color, as indicated. For s-wave annihilation (left panel) we see that the FIRE runs sometime produce higher J-factors (up to a factor of $\sim 6$) at small angle and sometimes give decreased J-factors (as small as $\sim 0.3$ of the DMO value). Because for s-wave annihilation the J-factor depends only on the density, this behavior tracks what seen for the density profiles (Figure \ref{fig:ratios}).  Sometimes feedback has produced a cored-out central density profile, leading to a lower central J-factor; sometimes baryonic contraction is more important, and this creates higher central densities and higher J-factors at small angle.  

The middle and right panels of Figure \ref{fig:djratio} show that the J-factor ratios are higher, in all cases, for the FIRE runs for p-wave and d-wave annihilation.  This is because the dark matter velocities are always enhanced (see Figure \ref{fig:profiles}) by factors of $2.5-4$, which is enough to boost the p-wave and d-wave annihilation with respect to DMO runs, even in the cases where the central density is slightly smaller.  Typical amplification factors are $\sim 10$ for p-wave and $\sim 100$ d-wave.   In cases like \texttt{M12i} and \texttt{Romulus}, where the dark matter density is also higher in the FIRE runs, the p-wave and d-wave ratios can very large (factors of $\sim 40-60$ and $\sim 400-500$, respectively).  

Another way to see the difference between DMO and FIRE models is to compare the cumulative J-factors within  annular angle $ \psi $ from the Galactic Center.  
Figure \ref{fig:Jtot} shows this quantity for each DMO (left) and FIRE (right) run.  Each halo pair has a unique color, as indicated. For s-wave annihilation (top left and right panels) we see that the FIRE runs generally do have larger integrated J-factors within 20 degrees of the Galactic Center, even in cases (like \texttt{Thelma}) that have somewhat smaller signals within 10 degrees.  
The middle and bottom panels of Figure \ref{fig:Jtot} show that the J-factor cumulative totals are higher, in all cases, for the FIRE runs for p-wave and d-wave annihilation.  Though the cumulative totals are more enhanced within $\sim 10$ degrees owing the shape of the profile.

In the next section we briefly explore implications of our results for dark matter indirect detection.

\section{Implications}
\label{sec:implications}

One of our primary results is that the DM velocities in our full galaxy formation runs are significantly higher than would be expected from DMO runs; this elevates the expected signal for fixed cross section in p-wave and d-wave models (see, e.g. Figure \ref{fig:Jtot}). In what follows we aim to provide a schematic illustration of how our results may impact attempts to constrain dark matter models with thermal abundance cross sections, especially those with velocity dependence. We use the published results of \citet{Abazajian20} in this illustrative example, and adopt a simple scaling of their published limits to provide a first-order sketch of how our results may impact future attempts to constrain s, p, and d-wave annihilation.

As discussed in the introduction, the realization that the observed Galactic Center gamma-ray excess has a non-circular, boxy shape that traces the Galactic Bulge \citep{Macias18,Bartels18} has allowed 
\citet{Abazajian20} to rule out a number of thermal-abundance WIMP models with s-wave annihilation channels for $m_\chi \lesssim 500$ GeV. Such a result motivates the exploration of p-wave (and d-wave models).  There is a general expectation that velocity-suppressed p-wave and d-wave annihilation will be far from detectable in Milky Way  \citep[though see][]{johnson2019search}. This is because typical DM velocities in the Galactic Center are usually thought to be $v \sim 100$ km s$^{-1}$ (based on DMO simulations), compared values $\sim 10^{3}$ times higher during thermal freeze out.  While, from the point of view of a model builder, such a suppression is ``good" because it evades direct-detection bounds,  from the point of view of an observer or experimentalist, this level of suppression is a potential nightmare: how can we detect such a signal?

The thick black lines in the upper set of panels Figure \ref{fig:constraints} reproduces the s-wave constraints published by \citet{Abazajian20}. The horizontal axis shows the WIMP mass and vertical axis is the velocity-averaged cross section. In our generalized language, the vertical axis specifically corresponds to the normalization $[\sigma v]_0$, defined by $\langle \sigma v \rangle = [\sigma v]_0 Q(v_{\rm rel})$ in Equation \ref{eq:sigma}, where $Q(v)=1$ is the s-wave case. Cross sections above the black line are excluded.  In deriving this constraint, they assumed a $b\bar{b}$ annihilation channel and a plausible range of Milky Way dark matter profiles (their "NFW" case) as expected from DMO simulations.    The dashed line shows the required cross section to produce the correct thermal abundance of dark matter observed \citep{Steigman12}.

The blue lines in Figure \ref{fig:constraints} provide  schematic estimates for how the \citet{Abazajian20} limit would shift for s- (top), p- (middle), and d-wave (bottom) annihilation for halos that match our simulation results. Here we have made the simplistic assumption that limit will scale in direct proportion to the integrated J-factor within $10^\circ$ of the Galactic Center.  The range of NFW profiles considered in \citet{Abazajian20} have central densities quite similar to our own \texttt{M12wDMO} case, and we use this to set the reference J-factor for the constraint: $J_s(<10^\circ) \equiv J_{\rm ref} =  1.7 \times 10^{22}$ GeV$^2$ cm$^{-3}$. Note that this reference J$_s$-factor is at the lower range of those from our DMO halos (See Table \ref{tab:one}). This is mainly due to the fact that while we normalized each halo to a local, solar dark matter density of $0.38$ GeV cm$^{-3}$, they assumed a median normalization of $0.28$ GeV cm$^{-3}$ at the solar radius.  Some of our DMO halos also deviate somewhat from a strict NFW shape, effectively making them slightly denser at $\sim 1$ kpc than that shape would predict for a fixed solar normalization \citep[e.g.][]{lazar2020dark}.

 Each blue line is a scaled version of the black line.  Specifically, for each halo in our suite, we determine the ratio of the reference J-factor from Abazajian et al. to the measured J-factor for s, p, and d-wave cases: $J_{\rm ref}/J_q(<10^\circ)$ (for $q=$ s, p, and d), and multiply the Abazajian limit by that ratio to estimate the implied limit.  

The top pair of panels shows how the implied limit scales for each of our DMO runs (left) and FIRE runs (right) in the s-wave case.  As mentioned above, even in the DMO case, our halos tend to have larger J-factors than the halos used in \citet{Abazajian20} because of our chosen local-density normalization.  The spread in lines comes about because of the halo-to-halo scatter. Interestingly, for the FIRE cases, which we regard as more realistic, the limit lies above all lines, though is within $\sim 25\%$ of the upper envelope.   One way to interpret this is that the Abazajian limit is conservative in comparison to our expectations, but not exceedingly so, especially considering the sensitivity to local dark matter density normalization.   In this sense, our results are unlikely to affect current constraints for the s-wave cross section significantly.  

The middle panels show the implied constraints for the case of p-wave annihilation, with DMO halos on the left and FIRE halos on the right.  The dotted line shows the required cross section normalization for the thermal abundance in the p-wave annihilation case.  The value is about $50$ times higher than the s-wave thermal abundance normalization to make up for the fact that the total cross section is suppressed by a factor $Q=(v/c)^2$ with $v/c \sim 0.15$ during freeze-out.  We work out the thermal abundance normalization explicitly for p-wave and d-wave dark matter in Appendix \ref{sec:appendix}. 

Note that the scaled limits  in the DMO p-wave panel are more than two orders of magnitude above the thermal cross section, which would suggest indirect detection for such a model is unlikely.  The right middle panel of Figure \ref{fig:constraints} unitizes what we believe to be more realistic p-wave J-factors from our FIRE runs.  The blue lines in this case suggest that a much more powerful constraint is possible for p-wave annihilation than would have been expected from DMO halos alone.  In particular, we see that for low-mass WIMPS, we may be within a factor of $\sim 10-20$ of detecting a p-wave annihilation signal if the Milky Way resembles halos like \texttt{Romulus}, \texttt{Juliet}, and \texttt{Romeo}, which have among our highest p-wave J-factors.

Finally, the bottom panels show the implied constraints for the case of d-wave annihilation.  The required cross section normalization for the thermal abundance (dotted line) in the d-wave annihilation case  is $\sim 2000$ times higher than the s-wave thermal abundance normalization (Appendix \ref{sec:appendix}), though it is still orders of magnitude below any of the scaled constraints.  Inferred constraints from the DMO halos (left) are five to six orders of magnitude above the thermal abundance normalization.  For the FIRE cases, the situation is slightly better (roughly four orders of magnitude) though still far out of reach.   

Of course, realistic constraints will require a careful analysis of Fermi-LAT Galactic Center observations, including templates for the stellar galactic and nuclear bulges, variations in the Galactic diffuse emission models, and a careful consideration of the shape of p-wave J-factor models of the kind show in Figures \ref{fig:map_juliet} and \ref{fig:map_m12c}, which tend to be even less boxy in the hydrodynamic runs than would be expected in DMO.   Based on the rough estimates presented here, such an analysis is certainly warranted. 

\section{Comparison to previous work}

As discussed in the introduction,  \citet{Board21} have presented an analysis similar to ours, comparing velocity-dependent J-factors in both DMO and hydrodynamic simulations. In this short subsection we compare our results to theirs.  Importantly, we are in general agreement on the key result: both of us find that the (more realistic) hydrodynamic simulations predict higher central dark matter particle velocities towards the Galactic center, and this enhances J-factors for p- and d-wave annihilation compared to DMO simulations.

For a more specific comparison, we must take into account three key differences in our efforts.  First, their work relies on lower-resolution simulations from the APOSTLE and Auriga suites.  This means that they can only make predictions to within 10 (APOSTLE) and 7 (Auriga) degrees, compared to our $2.75$ degree resolution. For this reason, we will make comparisons at 10 degrees in the discussion that follows. A second difference in our analysis is that we have re-scaled all of our simulations to have the same local dark matter density at mock solar locations, and have chosen a value that matches observational constraints for the Milky Way.  \citet{Board21} present J-factors based on raw simulation results, which will naturally scale as the local density squared by definition.   A third difference is that \citet{Board21} assume their halos are spherically symmetric in characterizing velocity moments, while we use direct particle counts to estimate the local density and velocity moments.  This allows us to place the observer in the plane of the galactic disk in constructing emission maps and also allows us explore how galaxy formation affects the shape of expected emission on the sky (we find it makes it much more round than naively would be expected).

The s-wave differential J-factor predictions for the DMO runs in \citet{Board21} are roughly $\sim 4$ times lower at 10 degrees than those shown in Figure \ref{fig:djdo} for our DMO halos.  This stems from the fact that their dark matter halos are less dense by a factor of $\sim 2$ at the solar location than our chosen normalization. Their hydrodynamic runs produce similar s-wave J-factors to our FIRE runs at 10 degrees ($\sim 50 \%$ lower in the median).  This is because their halos have become more dense in the center as a result of galaxy formation, bringing them closer, though still slightly below, the normalization we have chosen.  Note that, like \citet{Board21}, we also find that galaxy formation usually makes halos slightly denser, as can be seen in our un-normalized profiles shown in the upper left panel of Figure \ref{fig:profiles}.  Their DMO J-factors for p-wave and d-wave models are also consistent with ours modulo the density normalization factor.

Among the most important results in both of our papers is the systematic enhancement in p- and d-wave J-factors in hydrodynamic simulations compared to DMO cases.  While \citet{Board21} do not show ratios for each halo individually, they generally find that their DMO halos have p-wave J-factors at 10 degrees that are $\sim 5$ times lower than their full-physics runs at the same angular scale.  This is close to the typical ratio we present in Figure \ref{fig:djratio} at 10 degrees.  Similarly, for d-wave, their galaxy formation simulations have larger J-factors by $\sim 30$, and this is consistent with our typical ratio at 10 degrees as well.

As mentioned above, a key advance in the present paper is in our ability to push predictions towards the inner few degrees of the Galactic center, which is an important region for the gamma ray excesses seen in the Milky Way \citep[e.g.][]{karwin2017dark}. In these inner regions, we find that our FIRE halos show enhancements by factors as high as $\sim 50$ (p-wave) and $\sim 300$ (d-wave) -- both of which are $\sim 10$ times higher than the amplification seen 10 degrees in both our work (Figure \ref{fig:djratio}) and \citet{Board21}.

\section{Conclusions}

We have explored how galaxy formation affects predictions for the astrophysical J-factors of Milky Way size dark matter halos.  For a fixed particle physics model, astrophysical J-factors are directly proportional to the expected flux of Standard Model particles sourced by dark matter annihilation, and therefore provide a crucial input for  dark-matter indirect detection searches in the Milky Way (see Eq. \ref{eq:Jfactor_def}). In particular, we have used twelve FIRE zoom simulations of Milky Way-type galaxies along with dark-matter-only (DMO) versions of the same halos and worked out implications for both velocity-independent (s-wave) and velocity-dependent (p-wave and d-wave) annihilation cross sections.   

One significant result is that the central dark matter velocity dispersion in FIRE halos is {\em systematically} amplified by factors of $\sim 2.5-4$ compared to their DMO counterparts (right panel of Figure \ref{fig:ratios}).  The effect of galaxy-formation on the central dark matter density in the same halos is less systematic, sometimes increasing and sometimes decreasing the central density, with ratios ranging from $\sim 0.3-2.5$  (left panel of Figure \ref{fig:ratios}). For p-wave ($\propto (v/c)^2)$ and d-wave ($\propto (v/c)^4$) models, our FIRE-derived J-factors are amplified by factors of $\sim 3-60$ and $\sim 10-500$ compared to DMO runs (see Figure \ref{fig:djratio}).  FIRE halos generally produce J-factor profiles that are flatter (less peaked) towards the Galactic Center (see Figure \ref{fig:djdo}) and rounder on the sky (see images \ref{fig:map_juliet} and \ref{fig:map_m12c}).  Note that these differences occur despite the fact that we have normalized all of our halos to have the same local (solar location) dark matter density.  That is, these results are driven by differences in the {\em shape} of the underlying dark matter density and velocity dispersion profiles brought about by galaxy formation processes.

One basic implication of our results is that we expect p-wave and d-wave dark matter annihilation to produce more easily detectable signals than would have been expected from DMO halos.  For example, while it is typical to suspect that p-wave annihilation ($\propto (v/c)^2)$) is suppressed to undetectable levels~\footnote{Though see \citet{johnson2019search}, who have investigated the role of the central black hole in altering the dark matter velocity dispersion.} in the Milky Way today (where $v \ll c$), we showed in section \ref{sec:implications} that this may not be the case.  With the amplified velocities we see in our FIRE runs, the detection of (or interesting constraints on) thermal-relic p-wave dark matter may not be too far out of reach. In particular, by scaling the s-wave constraints from Fermi-LAT derived by \citet{Abazajian20},  we showed that a similar analysis could bring p-wave constraints to within a factor of $\sim 10$ of the naive thermal cross section (see the right middle panel of Figure \ref{fig:constraints}).  Future  analyses that include detailed simulation-inspired priors on the shape of the annihilation signal in these models, could potentially approach the thermal value.

Another result worth highlighting is that we see significant scatter in the density profiles (and associated s-wave J-factor profiles) in our FIRE runs.  As we begin to explore joint-constraints from multiple galaxies (e.g. M31 and the Milky Way) it will be important to allow for the possibility that even halos with similar halo masses are expected to have a large scatter in J-factor normalization.  Figure \ref{fig:djdo}, for example, shows that the variance in our FIRE s-wave J-factors is larger than one order of magnitude for our sample at few degrees from the Galactic Center, despite the fact that our halos have been fixed to have the same local density of dark matter at the solar location.  Similarly, the p- and d-wave J-factor enhancement in our FIRE runs can vary considerably from halo to halo. This finding for galaxies with similar properties could be important for future surveys looking for J-factor signals in both our own galaxy and beyond.

As a final point, we mention that the standard FIRE-2 implementation used in our simulations does not include the effect of AGN feedback on the gas distribution, which could affect central dark matter densities.  \citet{peirani2017density} have used the Horizon simulations to explore the effect of AGN feedback on the density profiles of dark matter halos in simulations with and without AGN. Their results show AGN feedback can reduce the central dark matter density compared to runs without, especially at early times and in higher mass halos.  However, at the Milky Way mass scale and at z=0, dark matter profiles of halos with and without AGN are very similar down to the radii that are converged in their simulations ($5$ kpc, their Figure 2).  This might suggest we would see very little difference for the Milky Way itself.  Nevertheless,  it is possible that at smaller radii ($\sim 400$ pc, of relevance for our work) there could be some reduction in density, even for a black hole as low mass as Sgr A*, which at $\sim 4 \times 10^6$M$_\odot$ \citep{Boehle16} is lower than expected from the tight correlation seen in ellipticals and galaxies with bulges \citep{Kormendy13}. Importantly,  we do not expect the enhancement in dark matter velocities to be diminished in runs with AGN feedback.  If anything, we might expect an even greater enhancement in dark matter velocities if AGN feedback were strong enough to affect the dark matter density. Indeed, in the vicinity of the black hole, it is possible that the velocity spike could be significant, further motivating continued studies of gamma-ray emission near the Galactic center \citep{johnson2019search}.

\section*{Acknowledgements}
We dedicate this paper to the memory of our dear friend and colleague José Antonio Florez Velázquez. We would like to thank Tyler Kelly for  helpful suggestions and assistance in the analysis.  We thank Kev Abazajian and Louis Strigari for useful discussions and we would like to thank the anonymous referee for their insightful suggestions that improved the quality of this work.

This work made use of the FIRE Studio software package \citep{FIRESTUDIO}.

DM, JSB, and FJM were supported by NSF grant AST-1910346.  
ZH is supported by a Gary A. McCue postdoctoral fellowship at UC Irvine.
PH is supported by  NSF Research Grants 1911233 \&\ 20009234, NSF CAREER grant 1455342, NASA grants 80NSSC18K0562, HST-AR-15800.001-A. Numerical calculations were run on the Caltech compute cluster ``Wheeler,'' allocations FTA-Hopkins/AST20016 supported by the NSF and TACC, and NASA HEC SMD-16-7592.

AW received support from: NSF grants CAREER 2045928 and 2107772; NASA ATP grant 80NSSC20K0513; HST grants AR-15809 and GO-15902 from STScI; a Scialog Award from the Heising-Simons Foundation; and a Hellman Fellowship.

MBK acknowledges support from NSF CAREER award AST-1752913, NSF grant AST-1910346, NASA grant NNX17AG29G, and HST-AR-15006, HST-AR-15809, HST-GO-15658, HST-GO-15901, HST-GO-15902, HST-AR-16159, and HST-GO-16226 from the Space Telescope Science Institute, which is operated by AURA, Inc., under NASA contract NAS5-26555.

\section*{Data Availability}

The data supporting the plots within this article are available on reasonable request to the corresponding author. A public version of the GIZMO code is available at \href{http://www.tapir.caltech.edu/~phopkins/Site/GIZMO.html}{http://www.tapir.caltech.edu/~phopkins/Site/GIZMO.html}.



\bibliographystyle{mnras}
\bibliography{Velocity-Dependent-J-Factor.bib}




\appendix

\section{Thermal cross section for velocity-dependent annihilation}
\label{sec:appendix}

In this subsection we provide an estimate for the required thermal cross section to match the observed dark matter abundance for models with a velocity-dependent annihilation cross section. We are assuming that 
\begin{equation}
    \langle \sigma v \rangle = [\sigma v]_0 Q(v_{\rm rel})
    \label{eq:sigma}
\end{equation}
where $[\sigma v]_0$ is a normalization constant  and $Q(v) = 1$, $v^2/c^2$, and $v^4/c^4$, for s, p, and d-wave annihilation, respectively. For the standard s-wave case ($Q=1$), the normalization required for thermal abundance was worked out carefully by \citet{Steigman12}, who find
\begin{equation}
    [\sigma v]_0^{T \, s} \simeq 2.3 \times 10^{-26} \rm{cm}^3 \, {\rm s}^{-1}
    \label{eq:Steigman}
\end{equation}
for WIMP masses above $\sim 10$ GeV \citep{Steigman12}.
Our goal here is to determine how the required normalization changes for p-wave and d-wave. 

We will follow the textbook treatment of WIMP freeze-out \citep[see, e.g.][]{kolb1990,dodelson2003modern,mo2010galaxy,Lisanti17}, with the standard assumption that WIMPs are their own anti-particle and initially in thermal equilibrium with Standard Model particles in the early universe with photon temperature $T$.  In this case the dark matter number density $n$ can be tracked using a simplified version of the Boltzmann equation 
\begin{equation}
  \frac{d n }{dt } + 3 H(t) n =  \langle \sigma v \rangle (n_{eq}^2 - n^2),
 \label{eq:Boltz}
\end{equation} 
where $H(t)$ is the Hubble parameter. Here, $n_{eq}$ is the equilibrium number density in physical coordinates \citep{kolb1990}, which scales with the expansion factor $a$ as $n_{eq} \propto T^3 \propto a^{-3}$ for relativistic species and becomes thermally suppressed for non-relativistic species $\propto e^{-m/T}$, where $m$ is the dark matter particle mass. It is useful to rewrite this equation in terms of a new variable $Y \equiv n/T^3$, which scales out the expansion of the universe and eliminates the $3H(t)$ term on the left-hand side of Eq. \ref{eq:Boltz}:  
\begin{equation}
    \frac{d Y}{dt} = T^3 \langle \sigma v \rangle (Y_{eq}^2 - Y^2) \: \longrightarrow \: \frac{d Y}{d x} = \frac{\lambda(x)}{x^2} (Y_{eq}^2 - Y^2),
    \label{eq:Boltz2}
\end{equation}
where $Y_{eq} \equiv n_{eq}/T^3$ and the arrow points to an equivalent equation that uses a new time variable $x = m/T$.  The variable $\lambda(x)  \equiv m^3 \langle \sigma v \rangle/H(m)$, where $H(m)$ in the denominator is the Hubble parameter evaluated at the time when the temperature $T=m$.  In the standard (s-wave) treatment, $\lambda$ is a constant. More generally  $\lambda(x) \propto [\sigma v]_0 \, x^{-n}$ with $n=0$, $1$, and $2$ for s-wave, p-wave, and d-wave,  respectively.

While there is no analytic solution to Equation \ref{eq:Boltz2}, we can estimate how the ultimate abundance will scale with input parameters.   At early times ($x \ll 1$), the coefficient $\lambda/x^2$ will be very large and $Y$ will track the equilibrium abundance.  As time progresses, the $\lambda/x^2$ term will become smaller than unity as the annihilation rate drops below the expansion rate.  After this point, the dark matter particles will no longer track the equilibrium abundance but instead ``freeze out".  For typical models, the freeze-out time is well into the non-relativistic regime $x_f \simeq 25$.  After this time ($x \gtrsim x_f$), we expect $Y \gg Y_{eq}$ because $Y_{eq} \propto \exp(-x)$.  In this limit, Equation \ref{eq:Boltz2} simplifies to 
\begin{equation}
    \frac{d Y}{d x} = \frac{- \lambda(x)}{x^{2}} Y^2.
\end{equation}
We can estimate $Y$ today by solving the above equation as an integral over $x$ from the freeze-out time $x_f$ to today $x_0 = m/T_0 \simeq \infty$.  The value of $Y$ today, $Y_\infty$, maps directly to the dark matter density today: $\Omega_{dm} \propto m Y_{\infty}/\rho_{\rm crit}$.  This allows us to write 
\begin{equation}
    \Omega_{dm} \propto  \left[\int_{x_f}^{\infty} \frac{[\sigma v]_0}{x^{n+2}}dx \right]^{-1} =  \frac{(n+1)x_f^{n+1}}{[\sigma v]_0}.
    \label{eq:omega_freeze}
\end{equation}
Equation \ref{eq:omega_freeze} allows us to estimate how $[\sigma v]_0$ must change as a function of velocity-dependence $n$ in order to keep $\Omega_{\rm dm}$ fixed:
\begin{equation}
    [\sigma v]_0 \propto (n+1) x_f^{n+1}.
    \label{eq:scaling}
\end{equation}

Since we expect $x_f$ to vary only logarithmically with the cross section, let us assume $x_f \simeq 25$ in all cases.  According to Equation \ref{eq:scaling},  scaling the s-wave cross section normalization (\ref{eq:Steigman}) by a factor $2 x_f^2$ will give us a thermal cross section estimate for p-wave ($n=1$):
\begin{equation}
    [\sigma v]_0^{T \, p} \approx 50 [\sigma v]_0^{T \, s} = 1.2 \times 10^{-24} \rm{cm}^3 \, {\rm s}^{-1}.
\end{equation} 
For d-wave ($n=2$), the scaling is $3 x_f^3$ and we have
\begin{equation}
  [\sigma v]_0^{T \, d} \approx 1875 [\sigma v]_0^{T \, s}  = 4.3 \times 10^{-23} \rm{cm}^3 \, {\rm s}^{-1}.
\end{equation} 
We have checked the above two numbers by solving Equation \ref{eq:Boltz2} numerically and find them to be good approximations.

\section{Comparison of baryonic matter densities to dark matter in the Galactic Centers of FIRE -2  halos }

\begin{figure}

						\includegraphics[height =0.9\columnwidth, trim = 0 0 0 0 ]{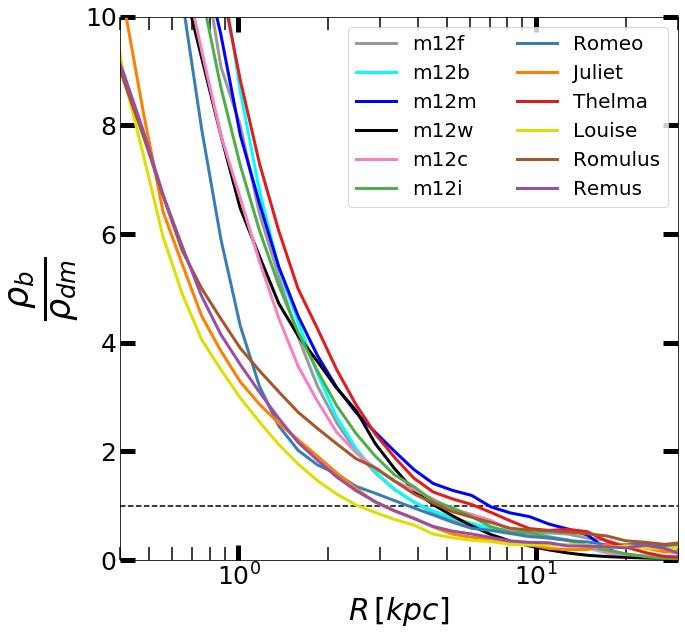}
    \caption{Ratio of baryonic density to dark matter density at as a function of radius  within FIRE halos. Note all halos are baryon dominated at inner radii until falling to roughly 1:1 near a mock solar location $\sim 8$ kpc.  }
    \label{fig:ratio}
\end{figure}

We examined the ratios of the densities of ordinary matter (gas and stars) to that of dark matter in regions near the galactic center of FIRE -2 halos. Our results are shown in Figure \ref{fig:ratio}. As anticipated, the halos are dominated by baryonic matter near their most central regions, and then fall below an even 1:1 at 3-8 kpc from the center.  Interestingly, the ratios vary substatially from halo to halo. For example, \texttt{Louise} and a few others have a baryonic content roughly 10 times larger than their dark matter content within the innermost resolved region (400 pc), but other halos, such as Thelma have ratios of up to 20 to 30 times greater at the same radius. Understanding the run-to-run variance in baryonic to dark matter density ratios will be a goal of future work.

\section{Comparison to Local Maxwellian Assumption}
\label{sec:maxwell}

\begin{figure}
			\includegraphics[width =0.95\columnwidth,trim = 0 0 0 0]{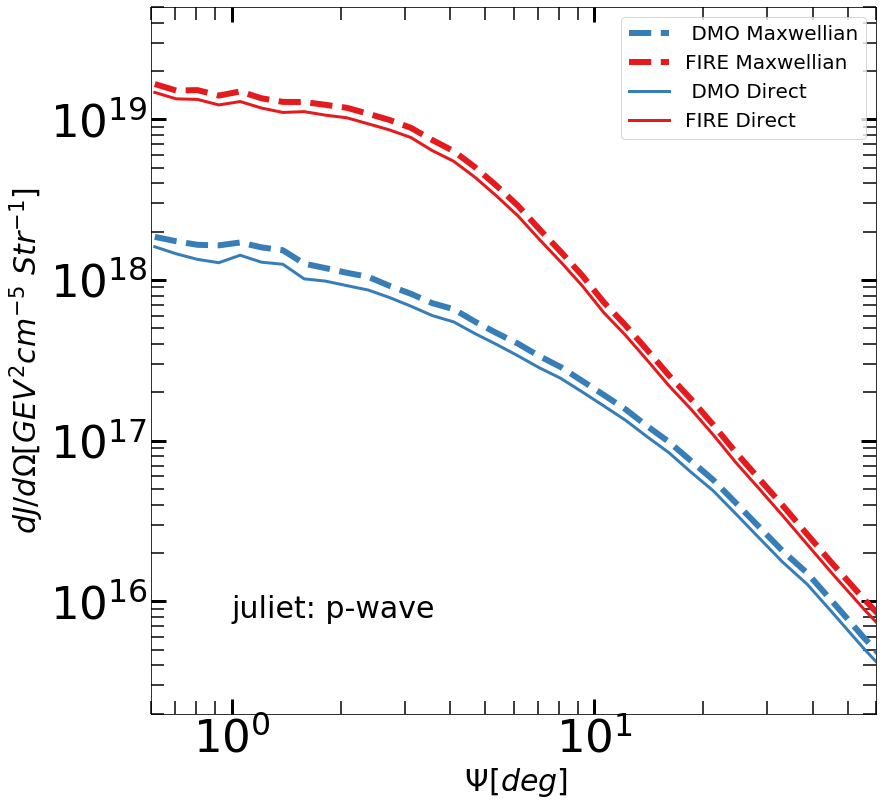} 
		\includegraphics[width =0.95\columnwidth, trim = 0 0 0 0]{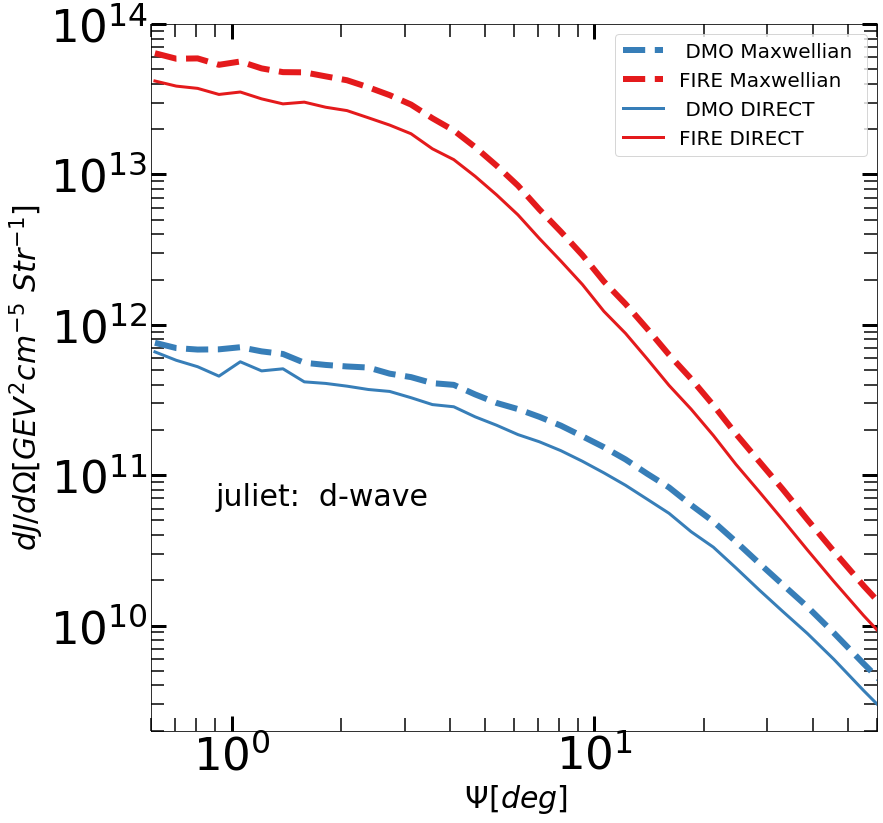} 
		
    \caption{Comparison of differential p-wave (left) and d-wave (right) J-factor for \texttt{Juliet} using a local spherical Maxwelllian approximation (dashed) and direct velocity moments (solid). We see that the direct calculation yields results that are $\sim 20\%$ lower for p-wave and $\sim 40\%$ lower for d-wave. Similar results were obtained for all the other halos. }
    \label{fig:djratio_comp}
\end{figure}

As discussed in Section \ref{sec:approach}, we calculate the p-wave and d-wave J-factors (Equations \ref{eqn:dp} and \ref{eqn:dd}) by measuring local relative velocity moments $\mu_2$ and $\mu_4$ in the vicinity of each dark matter particle in the simulation.  In this section we ask how our results would have changed had we instead estimated the velocity moments using the local velocity dispersion $ \sigma_v^2 = \langle v^2 \rangle$ and assumed a spherically-symmetric Maxwellian approximation. With this assumption, the second moment of the relative velocity would be $\mu_2 =   \langle (\vec{v_1} - \vec{v_2})^2 \rangle  = 2 \langle v^2 \rangle$, since the cross terms vanish under spherical symmetry.  The same assumptions give  $\mu_4 = \langle (\vec{v_1} - \vec{v_2})^4 \rangle  = 2 \langle v^4 \rangle + 2 \langle v^2 \rangle^2 = 48/9 \langle v^2 \rangle^2 $. Here we are using  $\langle v^4 \rangle = 15/9 \, \langle v^2 \rangle^2$ from the Maxwellian assumption and again assuming the cross terms vanish by spherical symmetry.
{Figure \ref{fig:djratio_comp} shows the comparison of our standard treatment (solid) to the spherical Maxwellian treatment (dashed) for \texttt{Juliet}.  We see that the direct calculation yields results that are $\sim 20\%$ lower for p-wave and $\sim 40\%$ lower for d-wave. Similar results were obtained for all the other halos.


\bsp	
\label{lastpage}
\end{document}